\documentclass[11pt,a4paper]{article}
\pdfoutput=1

\usepackage{jheppub}
\usepackage[utf8]{inputenc}
\usepackage[T1]{fontenc}
\usepackage{multirow}
\usepackage{ifthen}

\newcommand{\dd}{\mathrm{d}}
\newcommand{\eVq}{\ensuremath{\text{eV}^2}}
\newcommand{\Dmq}{\Delta m^2}
\newcommand{\Nuc}[2][]{{\ensuremath{\ifthenelse{\equal{#1}{}}{}{\mbox{}^{#1}}\text{#2}}}}

\makeatletter
\DeclareRobustCommand\recite[1]{\begingroup\@fileswfalse\cite{#1}\endgroup}
\makeatother

\graphicspath{{Figures/}}

\allowdisplaybreaks

\title{Status of Direct Determination of Solar Neutrino Fluxes after Borexino}

\author[a,b,c]{M.~C.~Gonzalez-Garcia,}
\affiliation[a]{Departament de F\'isica Qu\`antica i Astrof\'isica and
  Institut de Ci\`encies del Cosmos, Universitat de Barcelona,
  Diagonal 647, E-08028 Barcelona, Spain}
\affiliation[b]{Instituci\'o Catalana de Recerca i Estudis
  Avan\c{c}ats (ICREA), Pg.\ Lluis Companys 23, E-08010 Barcelona,
  Spain}
\affiliation[c]{C.N.~Yang Institute for Theoretical Physics, Stony
  Brook University, Stony Brook, NY 11794-3840, USA}
\emailAdd{maria.gonzalez-garcia@stonybrook.edu}

\author[d]{Michele Maltoni,}
\emailAdd{michele.maltoni@csic.es}
\affiliation[d]{Instituto de F\'isica Te\'orica (IFT-CFTMAT),
  CSIC-UAM, Calle de Nicol\'as Cabrera 13--15, Campus de Cantoblanco,
  E-28049 Madrid, Spain}

\author[a]{Jo\~ao Paulo Pinheiro,}
\emailAdd{joaopaulo.pinheiro@fqa.ub.edu}

\author[e,f]{Aldo M.~Serenelli}
\affiliation[e]{Institute of Space Sciences (ICE, CSIC), Campus UAB,
  Carrer de Can Magrans s/n, 08193 Cerdanyola del Valls, Spain}
\affiliation[f]{Institut d'Estudis Espacials de Catalunya (IEEC),
  C/Gran Capit\`a 2-4, 08034 Barcelona, Spain}
\emailAdd{aldos@ice.csic.es}

\abstract{We determine the solar neutrino fluxes from the global
  analysis of the most up-to-date terrestrial and solar neutrino data
  including the final results of the three phases of Borexino.  The
  analysis are performed in the framework of three-neutrino mixing
  with and without accounting for the solar luminosity constraint.  We
  discuss the independence of the results on the input from the
  Gallium experiments.  The determined fluxes are then compared with
  the predictions provided by the latest Standard Solar Models.  We
  quantify the dependence of the model comparison with the assumptions
  about the normalization of the solar neutrino fluxes produced in the
  CNO-cycle as well as on the particular set of fluxes employed for
  the model testing.}

\preprint{IFT-UAM/CSIC-24-3, YITP-SB-23-39}


\begin{document}

\maketitle

\section{Introduction}

At present we know without a doubt that $(a)$ the Sun produces its
energy as a whole by fusing protons into \Nuc[4]{He}, $(b)$ electron
neutrinos are by-products of these processes, and $(c)$ an established
fraction of such neutrinos have either muon or tau flavour by the time
they interact with the detectors on Earth.  These three facts,
nowadays well established, are the result of more than six decades of
research in theoretical and experimental astrophysics and particle
physics.

More specifically the fusion of protons into \Nuc[4]{He} is known to
proceed via two different mechanisms: the pp-chain and the
CNO-cycle~\cite{Bethe:1939bt, Bahcall:1989ks}.  In each of these two
mechanisms electron neutrinos are produced in a well-known subset of
reactions.  In particular, in the pp-chain five fusion reactions among
elements lighter than $A = 8$ produce neutrinos which are labeled by
the parent reaction as \Nuc{pp}, \Nuc[7]{Be}, \Nuc{pep}, \Nuc[8]{B},
and \Nuc{hep} neutrinos.  In the CNO-cycle the abundance of \Nuc{C}
and \Nuc{N} acts as a catalyst, and the \Nuc[13]{N} and \Nuc[15]{O}
beta decays provide the primary source of neutrinos, while \Nuc[17]{F}
beta decay produces a subdominant flux.  For each of these eight
processes the spectral energy shapes of the produced neutrinos is
known, but the calculation of the rate of neutrinos produced in each
reaction requires dedicated modeling of the Sun.

Over the years several Standard Solar Models (SSMs), able to describe
the properties of the Sun and its evolution after entering the main
sequence, have been constructed with increasing level of
refinement~\cite{Bahcall:1987jc, TurckChieze:1988tj, Bahcall:1992hn,
  Bahcall:1995bt, Bahcall:2000nu, Bahcall:2004pz, PenaGaray:2008qe,
  Serenelli:2011py, Vinyoles:2016djt}.  Such models are numerical
calculations calibrated to match present-day surface properties of the
Sun, and developed under the assumption that the Sun was initially
chemically homogeneous and that mass loss is negligible at all moments
during its evolution up to the present solar age $\tau_\odot =
4.57$~Gyr.  The calibration is done in order to satisfy the
constraints imposed by the current solar luminosity $L_\odot$, radius
$R_\odot$, and surface metal to hydrogen abundance ratio
$(Z/X)_\odot$.  Refinements introduced over the years include more
precise observational and experimental information about the input
parameters (such as nuclear reaction rates and the surface abundances
of different elements), more accurate calculations of constituent
quantities (such as radiative opacity and equation of state), the
inclusion of new physical effects (such as element diffusion), and the
development of faster computers and more precise stellar evolution
codes.

The detection of solar neutrinos, with their extremely small
interaction cross sections, enable us to see into the solar interior
and verify directly our understanding of the
Sun~\cite{Bahcall:1964gx}~-- provided, of course, that one counts with
an established model of the physics effects relevant to their
production, interaction, and propagation.  The Standard Model of
particle physics was thought to be such established framework, but it
badly failed at the first attempt of this task giving rise to the
so-called ``solar neutrino problem''~\cite{Bahcall:1968hc,
  Bahcall:1976zz}.  Fortunately we lay here, almost fifty years after
that first realization of the problem, with a different but well
established framework for the relevant effects in solar neutrino
propagation.  A framework in which the three flavour neutrinos
($\nu_e$, $\nu_\mu$, $\nu_\tau$) of the Standard Model are mixed
quantum superpositions of three massive states $\nu_i$ ($i=1,2,3$)
with different masses.  This allows for the flavour of the neutrino to
oscillate from production to detection~\cite{Pontecorvo:1967fh,
  Gribov:1968kq}, and for non-trivial effects (the so called
LMA-MSW~\cite{Wolfenstein:1977ue, Mikheev:1986gs} flavour transitions)
to take place when crossing dense regions of matter.  Furthermore, due
to the wealth of experiments exploring neutrino oscillations, the
value of the neutrino properties governing the propagation of solar
neutrinos, mass differences and mixing angles, are now precisely and
independently known.

Armed with this robust particle physics framework for neutrino
production, propagation, and detection, it is possible to turn to the
observation of solar neutrino experiments to test and refine the SSM.
Unfortunately soon after the particle physics side of the exercise was
clarified, the construction of the SSM run into a new puzzle: the so
called ``solar composition problem''.  In brief, SSMs built in the
1990's using the abundances of heavy elements on the surface of the
Sun from Ref.~\cite{Grevesse1998} (GS98) had notable successes in
predicting other observations, in particular helioseismology
measurements such as the radial distributions of sound speed and
density~\cite{Bahcall:1992hn, Bahcall:1995bt, Bahcall:2000nu,
  Bahcall:2004pz}.  But in the 2000's new determinations of these
abundances became available and pointed towards substantially lower
values, as summarized in Ref.~\cite{Asplund2009} (AGSS09).  The SSMs
built incorporating such lower metallicities failed at explaining the
helioseismic observations~\cite{Bahcall:2004yr}.
For almost two decades there was no successful solution of this puzzle
as changes in the modeling of the Sun did not seem able to account for
this discrepancy~\cite{Castro:2007, Guzik:2010, Serenelli:2011py}.
Consequently two different sets of SSMs were built, each based on the
corresponding set of solar abundances~\cite{Serenelli:2009yc,
  Serenelli:2011py, Vinyoles:2016djt}.

With this in mind, in Refs.~\cite{Gonzalez-Garcia:2009dpj,
  Bergstrom:2016cbh} we performed solar model independent analysis of
the solar and terrestrial neutrino data available at the time, in the
framework of three-neutrino masses and mixing, where the flavour
parameters and all the solar neutrino fluxes were simultaneously
determined with a minimum set of theoretical priors.  The results were
compared with the two variants of the SSM, but they were not precise
enough to provide a significant discrimination.

Since then there have been a number of developments.  First of all, a
substantial amount of relevant data has been accumulated, in
particular the full spectral information of the
Phase-II~\cite{Borexino:2017rsf} and Phase-III~\cite{BOREXINO:2022abl}
of Borexino and their results based on the correlated integrated
directionality (CID) method~\cite{Borexino:2023puw}.  All of them have
resulted into the first positive observation of the neutrino fluxes
produced in the CNO-cycle which are particularly relevant for
discrimination among the SSMs.

On the model front, an update of the AGSS09 results was recently
presented by the same group (AAG21)~\cite{Asplund2021}, though leading
only to a slight revision upwards of the solar metallicity.  Most
interestingly, almost simultaneously a new set of results
(MB22)~\cite{Magg:2022rxb}, based on similar methodologies and
techniques but with different atomic input data for the critical
oxygen lines among other differences, led to a substantial change in
solar elemental abundances with respect to AGSS09 (see the original
reference for details).  The outcome is a set of solar abundances
based on three-dimensional radiation hydrodynamic solar atmosphere
models and line formation treated under non-local thermodynamic
equilibrium that yields a total solar metallicity comparable to those
of the ``high-metallicity'' results by GS98.

Another issue which has come up in the interpretation of the solar
neutrino results is the appearance of the so-called ``gallium
anomaly''.  In brief, it accounts for the deficit of the rate of
events observed in Gallium source experiments with respect to the
expectation.  It was originally observed in the calibration of the
gallium solar-neutrino detectors GALLEX~\cite{GALLEX:1997lja,
  Kaether:2010ag} and SAGE~\cite{SAGE:1998fvr, Abdurashitov:2005tb}
with radioactive \Nuc[51]{Cr} and \Nuc[37]{Ar} sources, and it has
been recently confirmed by the BEST collaboration with a dedicated
source experiment using a \Nuc[51]{Cr} source with high statistical
significance~\cite{Barinov:2021asz, Barinov:2022wfh}.

The solution of this puzzle is an open question in neutrino physics
(see Ref.~\cite{Brdar:2023cms} and reference therein for a recent
discussion of --~mostly unsuccessful~-- attempts at explanations in
terms of standard and non-standard physics scenarios).  In particular,
in the framework of 3$\nu$ oscillations the attempts at explanation
(or at least alleviation) of the anomaly invoke the uncertainties of
the capture cross section~\cite{Berryman:2021yan, Giunti:2022xat,
  Elliott:2023xkb}.  With this motivation, in this work we have
studied the (in)sensitivity of our results to the intrinsic
uncertainty on the observed neutrino rates in the Gallium experiments
posed by possible modification of the capture cross section in
Gallium, or equivalently, of the detection efficiency of the Gallium
solar neutrino experiments.

All these developments motivate the new analysis which we present in
this paper with the following outline.  In Sec.~\ref{sec:frame} we
describe the assumptions and methodology followed in our study of the
neutrino data.  As mentioned above, this work builds upon our previous
solar flux determination in Refs.~\cite{Gonzalez-Garcia:2009dpj,
  Bergstrom:2016cbh}.  Thus in Sec.~\ref{sec:frame}, for convenience,
we summarize the most prominent elements common to those analyses, but
most importantly, we detail the relevant points in which the present
analysis method deviates from them.  The new determination of the
solar fluxes is presented in Sec.~\ref{sec:res} where we also discuss
and quantify the role of the Gallium experiments and address their
robustness with respect to the Gallium anomaly.  In Sec.~\ref{sec:CNO}
we have a closer look at the determination of the neutrino fluxes from
the CNO-cycle and its dependence on the assumptions on the relative
normalization of the fluxes produced in the three relevant reactions.
In Sec.~\ref{sec:compaSSM} we compare our determined fluxes with the
predictions of the SSMs in the form of a \emph{parameter goodness of
fit} test, and quantify the output of the test for the assumptions in
the analysis.  We summarize our findings in Sec.~\ref{sec:summary}.
The article is supplemented with a detailed Appendix~\ref{app:borex}
in which we document our analysis of the Borexino III spectral data
(and also their recent analysis employing the CID method).

\section{Analysis framework}
\label{sec:frame}

In the analysis of solar neutrino experiments we include the total
rates from the radiochemical experiments
Chlorine~\cite{Cleveland:1998nv}, Gallex/GNO~\cite{Kaether:2010ag},
and SAGE~\cite{Abdurashitov:2009tn}, the spectral and day-night data
from the four phases of Super-Kamiokande~\cite{Hosaka:2005um,
  Cravens:2008aa, Abe:2010hy, SK:nu2020}, the results of the three
phases of SNO in terms of the parametrization given in their combined
analysis~\cite{Aharmim:2011vm}, and the full spectra from Borexino
Phase-I~\cite{Bellini:2011rx}, Phase-II~\cite{Borexino:2017rsf}, and
Phase-III~\cite{BOREXINO:2022abl}, together with their latest results
based on the correlated integrated directionality (CID)
method~\cite{Borexino:2023puw}.  Details of our Borexino Phase-III and
CID data analysis, which is totally novel in this article, are
presented in Appendix~\ref{app:borex}.

In the framework of three neutrino masses and mixing the expected
values for these solar neutrino observables depend on the parameters
$\Dmq_{21}$, $\theta_{12}$, and $\theta_{13}$ as well as on the
normalizations of the eight solar fluxes.  Thus besides solar
experiments, we also include in the analysis the separate DS1, DS2,
DS3 spectra from KamLAND~\cite{Gando:2013nba} which in the framework
of three neutrino mixing also yield information on the parameters
$\Dmq_{21}$, $\theta_{12}$, and $\theta_{13}$.

In what follows we will use as normalization parameters for the solar
fluxes the reduced quantities:
\begin{equation}
  \label{eq:redflux}
  f_i = \frac{\Phi_i}{\Phi_i^\text{ref}}
\end{equation}
with $i = \Nuc{pp}$, \Nuc[7]{Be}, \Nuc{pep}, \Nuc[13]{N}, \Nuc[15]{O},
\Nuc[17]{F}, \Nuc[8]{B}, and \Nuc{hep}.  In this work the numerical
values of $\Phi_i^\text{ref}$ are set to the predictions of the latest
GS98 solar model, presented in Ref.~\cite{Magg:2022rxb, B23Fluxes}.
They are listed in Table~\ref{tab:lumcoef}.
The methodology of the analysis presented in this work builds upon our
previous solar flux determination in
Refs.~\cite{Gonzalez-Garcia:2009dpj, Bergstrom:2016cbh}, which we
briefly summarize here for convenience, but it also presents a number
of differences besides the additional data included as described next.

The theoretical predictions for the solar and KamLAND observables
depend on eleven parameters: the eight reduced solar fluxes
$f_{\Nuc{pp}}$, \dots, $f_{\Nuc{hep}}$, and the three relevant
oscillation parameters $\Dmq_{21}$, $\theta_{12}$, $\theta_{13}$.  In
our analysis we include as well the complementary information on
$\theta_{13}$ obtained after marginalizing over $\Dmq_{3\ell}$,
$\theta_{23}$ and $\delta_\text{CP}$ the results of all the other
oscillation experiments considered in NuFIT-5.2~\cite{nufit-5.2}.
This results into a prior $\sin^2\theta_{13} = 0.0223\pm 0.0006$,
\textit{i.e.}, $\theta_{13} = 8.59^\circ\, (1\pm 0.014)$.  Given the
weak dependence of the solar and KamLAND observables on $\theta_{13}$,
including such prior yields results which are indistinguishable from
just fixing the value of $\bar\theta_{13}=8.59^\circ$.

Throughout this work, we follow a frequentist approach in order to
determine the allowed confidence regions for these parameters (unlike
in our former works~\cite{Gonzalez-Garcia:2009dpj, Bergstrom:2016cbh}
where we used instead a Bayesian analysis to reconstruct their
posterior probability distribution function).  To this end we make use
of the experimental data from the various solar and KamLAND samples
($D_\text{solar}$ and $D_\text{KamLAND}$, respectively) as well as the
corresponding theoretical predictions (which depends on ten free
parameters, as explained above) to build the $\chi^2$ statistical
function
\begin{equation}
  \label{eq:chi2g}
  \chi^2_\text{global}(\vec\omega_\text{flux},\, \vec\omega_\text{osc})
  \equiv \chi^2_\text{solar}(D_\text{solar} \,|\,
  \vec\omega_\text{flux},\, \vec\omega_\text{osc})
  + \chi^2_\text{KamLAND}(D_\text{KamLAND}\,|\, \vec\omega_\text{osc}) \,,
\end{equation}
with $\vec\omega_\text{flux} \equiv (f_{\Nuc{pp}},\, \dots,\,
f_{\Nuc{hep}})$ and $\vec\omega_\text{osc} \equiv (\Dmq_{21},\,
\theta_{12},\, \bar\theta_{13})$.  In order to scan this
multidimensional parameter space efficiently, we make use of the
MultiNest~\cite{Feroz:2013hea, Feroz:2008xx} and
Diver~\cite{Martinez:2017lzg} algorithms.

\begin{table}\centering
  \catcode`?=\active\def?{\hphantom{0}}
  \begin{tabular}{r@{\hspace{20mm}}c@{\hspace{20mm}}c@{\hspace{20mm}}c}
    Flux & $\Phi_i^\text{ref}$ [$\text{cm}^{-2}\, \text{s}^{-1}$]
    & $\alpha_i$ [MeV] & $\beta_i$
    \\
    \hline
    \Nuc{pp}    & $5.960\times 10^{10}$ & $13.099?$ & $9.1864\times 10^{-1}$ \\
    \Nuc[7]{Be} & $4.854\times 10^{9?}$ & $12.552?$ & $7.1693\times 10^{-2}$ \\
    \Nuc{pep}   & $1.425\times 10^{8?}$ & $11.920?$ & $1.9987\times 10^{-3}$ \\
    \Nuc[13]{N} & $2.795\times 10^{8?}$ & $12.658?$ & $4.1630\times 10^{-3}$ \\
    \Nuc[15]{O} & $2.067\times 10^{8?}$ & $12.368?$ & $3.0082\times 10^{-3}$ \\
    \Nuc[17]{F} & $5.350\times 10^{6?}$ & $12.365?$ & $7.7841\times 10^{-5}$ \\
    \Nuc[8]{B}  & $5.025\times 10^{6?}$ & $?6.6305$ & $3.9205\times 10^{-5}$ \\
    \Nuc{hep}   & $7.950\times 10^{3?}$ & $?3.7355$ & $3.4944\times 10^{-8}$
  \end{tabular}
  \caption{The reference neutrino fluxes $\Phi_i^\text{ref}$ used for
    normalization (from Ref.~\recite{Magg:2022rxb, B23Fluxes}), the
    energy $\alpha_i$ provided to the star by nuclear fusion reactions
    associated with the $i^\text{th}$ neutrino flux (taken from
    Ref.~\recite{2021JPhG...48a5201V}), and the fractional
    contribution $\beta_i$ of the $i^\text{th}$ nuclear reaction to
    the total solar luminosity.}
  \label{tab:lumcoef}
\end{table}

The allowed range for the solar fluxes is further reduced when
imposing the so-called ``luminosity constraint'', \textit{i.e.}, the
requirement that the overall sum of the thermal energy generated
together with each solar neutrino flux coincides with the solar
luminosity~\cite{Spiro:1990vi}:
\begin{equation}
  \label{eq:lumsum1}
  \frac{L_\odot}{4\pi \, (\text{A.U.})^2}
  = \sum_{i=1}^8 \alpha_i \Phi_i \,.
\end{equation}
Here the constant $\alpha_i$ is the energy released into the star by
the nuclear fusion reactions associated with the $i^\text{th}$
neutrino flux; its numerical value is independent of details of the
solar model to an accuracy of one part in $10^4$ or
better~\cite{Bahcall:2001pf}.  A detailed derivation of this equation
and the numerical values of the coefficients $\alpha_i$ is presented
in Ref.~\cite{Bahcall:2001pf}, with some refinement and correction
following in~\cite{2021JPhG...48a5201V}.\footnote{We have explicitly
verified that the numerical differences between the results of the
analysis performed using the original $\alpha_i$ coefficients in
Ref.~\cite{Bahcall:2001pf} and those in
Ref.~\cite{2021JPhG...48a5201V} are below the quoted precision.}  The
coefficients employed in this work are listed in
Table~\ref{tab:lumcoef}.
In terms of the reduced fluxes Eq.~\eqref{eq:lumsum1} can be written
as:
\begin{equation}
  \label{eq:lumsum2}
  1 = \sum_{i=1}^8 \beta_i f_i
  \quad\text{with}\quad
  \beta_i \equiv
  \frac{\alpha_i \Phi_i^\text{ref}}{L_\odot \big/ [4\pi \, (\text{A.U.})^2]}
\end{equation}
where $\beta_i$ is the fractional contribution to the total solar
luminosity of the nuclear reactions responsible for the production of
the $\Phi_i^\text{ref}$ neutrino flux.  In
Refs.~\cite{Gonzalez-Garcia:2009dpj, Bergstrom:2016cbh} we adopted the
best-estimate value for the solar luminosity $L_\odot \big/ [4\pi \,
  (\text{A.U.})^2] = 8.5272 \times 10^{11}\, \text{MeV}\,
\text{cm}^{-2}\, \text{s}^{-1}$ given in Ref.~\cite{Bahcall:2001pf},
which was obtained from all the available satellite
data~\cite{Frohlich1998}.  This value was revised in
Ref.~\cite{kopp2011} using an updated catalog and calibration
methodology (see Ref.~\cite{Scafetta_2014} for a detailed comparative
discussion), yielding a slightly lower result which is now the
reference value listed by the PDG~\cite{Workman:2022ynf} and leads to
$L_\odot \big/ [4\pi \, (\text{A.U.})^2] = 8.4984 \times 10^{11} \,
\text{MeV} \, \text{cm}^{-2} \, \text{s}^{-1}$.  In this work we adopt
this new value when evaluating the $\beta_i$ coefficients listed in
Table~\ref{tab:lumcoef}.  Furthermore, in order to account for the
systematics in the extraction of the solar luminosity we now assign an
uncertainty of $0.34\%$ to the constraint in Eq.~\eqref{eq:lumsum2},
which we conservatively derive from the range of variation of the
estimates of $L_\odot$.
In what follows we will present results with and without imposing the
luminosity constraint.  For the analysis including the luminosity
constraint we add a prior
\begin{equation}
  \label{eq:priorLC}
  \chi^2_\text{LC}(\vec\omega_\text{flux})
  = \frac{\left( \displaystyle 1-\sum_{i=1}^8 \beta_i f_i\right)^2}{(0.0034)^2}
\end{equation}

Besides the imposition of the luminosity constraint in some of the
analysis, the flux normalizations are allowed to vary freely within a
set of physical constraints.  In particular:
\begin{itemize}
\item The fluxes must be positive:
  \begin{equation}
    \label{eq:fpos}
    \Phi_i \geq 0 \quad\Rightarrow\quad f_i \geq 0 \,.
  \end{equation}

\item Consistency in the pp-chain implies that the number of nuclear
  reactions terminating the pp-chain should not exceed the number of
  nuclear reactions which initiate it~\cite{Bahcall:1995rs,
    Bahcall:2001pf}:
  \begin{multline}
    \Phi_{\Nuc[7]{Be}} + \Phi_{\Nuc[8]{B}}
    \leq \Phi_{\Nuc{pp}} + \Phi_{\Nuc{pep}}
    \\
    \Rightarrow\quad
    8.12 \times 10^{-2} f_{\Nuc[7]{Be}}
    + 8.42 \times 10^{-5} f_{\Nuc[8]{B}}
    \leq f_{\Nuc{pp}} + 2.38 \times 10^{-3} f_{\Nuc{pep}} \,.
  \end{multline}

\item The ratio of the \Nuc{pep} neutrino flux to the \Nuc{pp}
  neutrino flux is fixed to high accuracy because they have the same
  nuclear matrix element.  We have constrained this ratio to match the
  average of the values in the five B23 SSMs
  (Sec.~\ref{sec:compaSSM}), with $1\sigma$ Gaussian uncertainty given
  by the difference between the values in the five models
  \begin{equation}
    \label{eq:pep-pp}
    \frac{f_{\Nuc{pep}}}{f_{\Nuc{pp}}} = 1.004 \pm 0.018 \,.
  \end{equation}
  Technically we implement this constraint by adding a Gaussian prior
  \begin{equation}
    \chi^2_\text{pep/pp}(f_{\Nuc{pp}},\, f_{\Nuc{pep}}) \equiv \left(
    \frac{f_{\Nuc{pep}} \big/ f_{\Nuc{pp}} - 1.004}{0.018}
    \right)^2.
  \end{equation}

\item For the CNO fluxes ($f\Phi_{\Nuc[13]{N}}$, $\Phi_{\Nuc[15]{O}}$,
  and $\Phi_{\Nuc[17]{F}}$) a minimum set of assumptions required by
  consistency are:
  \begin{itemize}
  \item The $\Nuc[14]{N}(p,\gamma) \Nuc[15]{O}$ reaction must be the
    slowest process in the main branch of the
    CNO-cycle~\cite{Bahcall:1995rs}:
    \begin{equation}
      \label{eq:CNOineq1}
      \Phi_{\Nuc[15]{O}} \leq \Phi_{\Nuc[13]{N}}
      \quad\Rightarrow\quad
      f_{\Nuc[15]{O}} \leq 1.35\, f_{\Nuc[13]{N}}
    \end{equation}

  \item the CNO-II branch must be subdominant:
    \begin{equation}
      \label{eq:CNOineq2}
      \Phi_{\Nuc[17]{F}} \leq \Phi_{\Nuc[15]{O}}
      \quad\Rightarrow\quad
      f_{\Nuc[17]{F}}\leq 40\, f_{\Nuc[15]{O}} \,.
    \end{equation}
  \end{itemize}
\end{itemize}
The conditions quoted above are all dictated by solar physics.
However, more practical reasons require that the CNO fluxes are
treated with a special care.  As discussed in detail in
Appendix~\ref{sec:bx3nfit}, the analysis of the Borexino Phase-III
spectra in Ref.~\cite{BOREXINO:2020aww} (which we closely reproduce)
has been optimized by the collaboration to maximize the sensitivity to
the overall CNO production rate, and therefore it may not be directly
applicable to a situation where the three \Nuc[13]{N}, \Nuc[15]{O} and
\Nuc[17]{F} flux normalizations are left totally free, subject only to
the conditions in Eqs.~\eqref{eq:CNOineq1} and~\eqref{eq:CNOineq2}.
Hence, following the approach of the Borexino collaboration in
Ref.~\cite{BOREXINO:2020aww}, we first perform an analysis where the
three CNO components are all scaled simultaneously by a unique
normalization parameter while their ratios are kept fixed as predicted
by the SSMs.  In order to avoid a bias towards one of the different
versions of the SSM we have constrained the two ratios to match the
average of the five B23 SSMs values
\begin{equation}
  \label{eq:CNOfix}
  \frac{\Phi_{\Nuc[15]{O}}}{\Phi_{\Nuc[13]{N}}} = 0.73
  \enspace\text{and}\enspace
  \frac{\Phi_{\Nuc[17]{F}}}{\Phi_{\Nuc[13]{N}}} = 0.016
  \quad\Rightarrow\quad
  \frac{f_{\Nuc[15]{O}}}{f_{\Nuc[13]{N}}} = 0.98
  \enspace\text{and}\enspace
  \frac{f_{\Nuc[17]{F}}}{f_{\Nuc[13]{N}}} = 0.85 \,.
\end{equation}
In these analysis, which we label <<CNO-Rfixed>>, the conditions in
Eq.~\eqref{eq:CNOfix} effectively reduces the number of free
parameters from ten to eight, namely the two oscillation parameters in
$\vec\omega_\text{osc}$ and six flux normalizations in
\begin{equation}
  \nonumber
  \vec\omega_\text{flux}^\text{CNO-Rfixed}
  \equiv (f_{\Nuc{pp}},\; f_{\Nuc[7]{Be}},\; f_{\Nuc{pep}},\; f_{\Nuc[13]{N}},\;
  f_{\Nuc[15]{O}} = 0.98\, f_{\Nuc[13]{N}},\;
  f_{\Nuc[17]{F}} = 0.85\, f_{\Nuc[13]{N}},\;
  f_{\Nuc[8]{B}},\; f_{\Nuc{hep}}) \,.
\end{equation}
In Sec.~\ref{sec:CNO} we will discuss and quantify the effect of
relaxing the condition of fixed CNO ratios.

\section{New determination of solar neutrino fluxes}
\label{sec:res}

\begin{figure}\centering
  \includegraphics[width=0.95\textwidth]{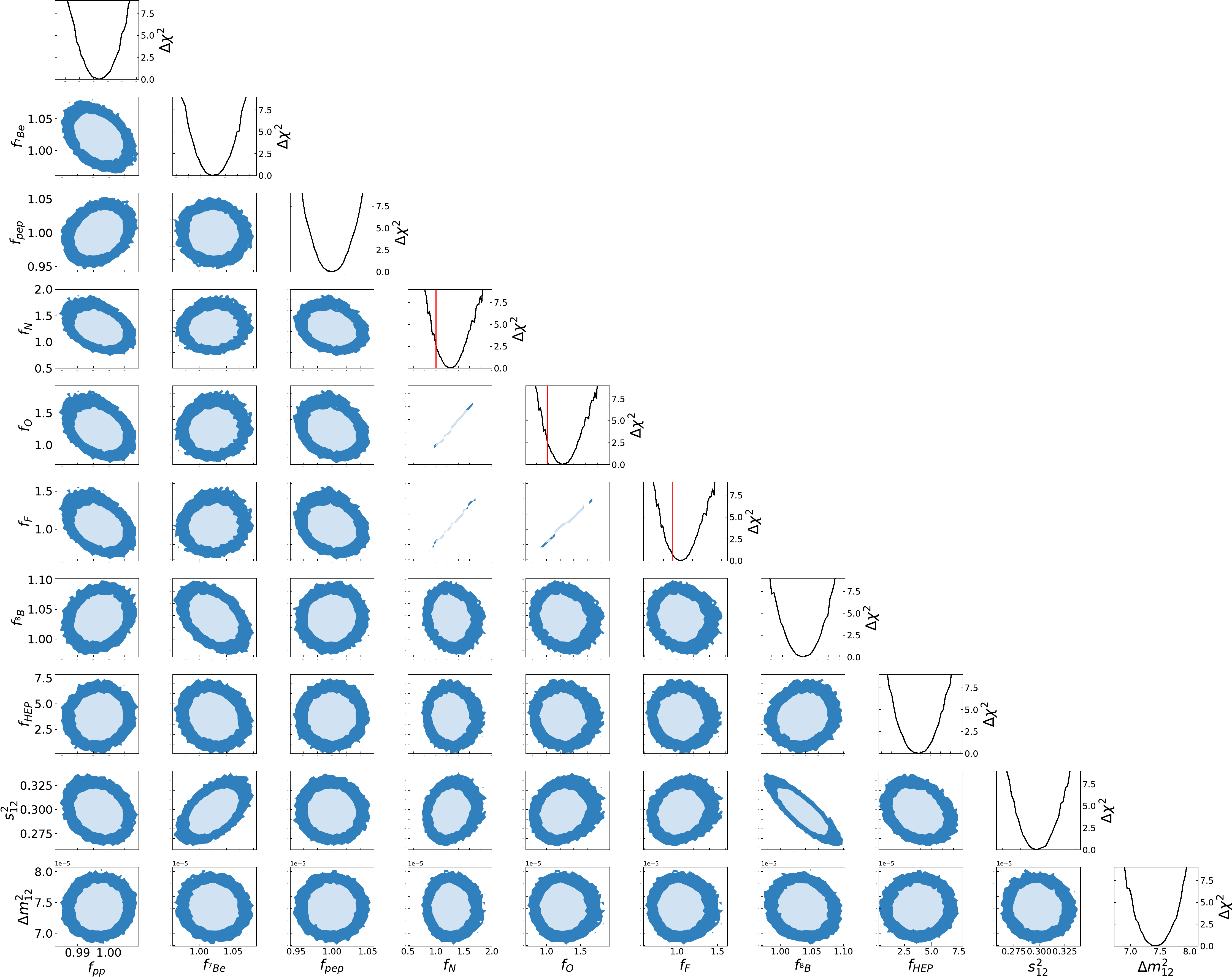}
  \caption{Constraints from our global analysis on the solar neutrino
    fluxes for the analysis with luminosity constraint and fixed
    ratios of the CNO fluxes (see Eq.~\eqref{eq:chi2wlccno}).  Each
    panel shows a two-dimensional projection of the allowed
    multidimensional parameter space after minimization with respect
    to the undisplayed parameters.  The regions correspond to 90\% and
    99\% CL (2 d.o.f.).  The curves in the rightmost panels show the
    marginalized one-dimensional $\Delta\chi^2_\text{wLC,CNO-Rfixed}$
    for each of the parameters.}
  \label{fig:triCNOLCGA1}
\end{figure}

We present first the results of our analysis with the luminosity
constraint and the ratios of the CNO fluxes fixed by the relations in
Eq.~\eqref{eq:CNOfix}, so that altogether for this case we construct
the $\chi^2$ function
\begin{equation}
  \label{eq:chi2wlccno}
  \chi^2_\text{wLC,CNO-Rfixed} \equiv
  \chi^2_\text{global}(\vec\omega_\text{osc},\, \vec\omega_\text{flux}^\text{CNO-Rfixed})
  + \chi^2_\text{pep/pp}(f_{\Nuc{pp}},\, f_{\Nuc{pep}})
  + \chi^2_\text{LC}(\vec\omega_\text{flux}^\text{CNO-Rfixed}) \,.
\end{equation}
The results of this analysis are displayed in
Fig.~\ref{fig:triCNOLCGA1}, where we show the two- and one-dimensional
projections of $\Delta\chi^2_\text{wLC,CNO-Rfixed}$.  From these
results one reads that the ranges at $1\sigma$ (and at the $99\%$ CL
in square brackets) for the two oscillation parameters are:
\begin{equation}
  \label{eq:bestosc}
  \begin{aligned}
    \Dmq_{21}
    &= 7.43_{-0.30}^{+0.30}\, [{}_{-0.49}^{+0.44}]
    \times 10^{-5} \eVq \,,
    \\[2pt]
    \sin^2\theta_{12}
    &= 0.300_{-0.017}^{+0.020}\, [{}_{-0.027}^{+0.031}] \,,
  \end{aligned}
\end{equation}
which are very similar to the results of NuFIT-5.2~\cite{nufit-5.2}
with the expected slight enlargement of the allowed ranges.  In other
words, within the $3\nu$ scenario the data is precise enough to
simultaneously constraint the oscillation parameters and the
normalizations of the solar flux components without resulting into a
substantial degradation of the former.
As for the solar fluxes, the corresponding ranges read:
\begin{equation}
  \label{eq:bestlc}
  \begin{aligned}
    f_{\Nuc{pp}}
    & = 0.9969_{-0.0039}^{+0.0041}\, [{}_{-0.0092}^{+0.0095}]
    \,, \qquad
    & \Phi_{\Nuc{pp}}
    & = 5.941_{-0.023}^{+0.024}\, [{}_{-0.055}^{+0.057}]
    \times 10^{10}~\text{cm}^{-2}~\text{s}^{-1} \,,
    \\
    f_{\Nuc[7]{Be}}
    & = 1.019_{-0.017}^{+0.020}\, [{}_{-0.041}^{+0.047}]
    \,, \qquad
    & \Phi_{\Nuc[7]{Be}}
    & = 4.93_{-0.08}^{+0.10}\, [{}_{-0.20}^{+0.23}]
    \times 10^{9}~\text{cm}^{-2}~\text{s}^{-1} \,,
    \\
    f_{\Nuc{pep}}
    & = 1.000_{-0.018}^{+0.016}\, [{}_{-0.042}^{+0.041}]
    \,, \qquad
    & \Phi_{\Nuc{pep}}
    & = 1.421_{-0.026}^{+0.023}\, [{}_{-0.060}^{+0.058}]
    \times 10^{8}~\text{cm}^{-2}~\text{s}^{-1} \,,
    \\
    f_{\Nuc[13]{N}}
    & = 1.25_{-0.14}^{+0.17}\, [{}_{-0.40}^{+0.47}]
    \,, \qquad
    & \Phi_{\Nuc[13]{N}}
    & = 3.48_{-0.40}^{+0.47}\, [{}_{-1.10}^{+1.30}]
    \times 10^{8}~\text{cm}^{-2}~\text{s}^{-1} \,,
    \\
    f_{\Nuc[15]{O}}
    & = 1.22_{-0.14}^{+0.17}\, [{}_{-0.39}^{+0.46}]
    \qquad
    & \Phi_{\Nuc[15]{O}}
    & = 2.53_{-0.29}^{+0.34}\, [{}_{-0.80}^{+0.94}]
    \times 10^{8}~\text{cm}^{-2}~\text{s}^{-1} \,,
    \\
    f_{\Nuc[17]{F}}
    & = 1.03_{-0.20}^{+0.20}\, [{}_{-0.48}^{+0.47}]
    \,, \qquad
    & \Phi_{\Nuc[17]{F}}
    & = 5.51_{-0.63}^{+0.75}\, [{}_{-1.75}^{+2.06}]
    \times 10^{7}~\text{cm}^{-2}~\text{s}^{-1} \,,
    \\
    f_{\Nuc[8]{B}}
    & = 1.036_{-0.020}^{+0.020}\, [{}_{-0.048}^{+0.047}]
    \,, \qquad
    & \Phi_{\Nuc[8]{B}}
    & = 5.20_{-0.10}^{+0.10}\, [{}_{-0.24}^{+0.24}]
    \times 10^{6}~\text{cm}^{-2}~\text{s}^{-1} \,,
    \\
    f_{\Nuc{hep}}
    & = 3.8_{-1.2}^{+1.1}\, [{}_{-2.7}^{+2.7}]
    \,, \qquad
    & \Phi_{\Nuc{hep}}
    & = 3.0_{-1.0}^{+0.9}\, [{}_{-2.1}^{+2.2}]
    \times 10^{4}~\text{cm}^{-2}~\text{s}^{-1} \,.
  \end{aligned}
\end{equation}
Notice that in Fig.~\ref{fig:triCNOLCGA1} we separately plot the
ranges for the three CNO flux normalization parameters, however they
are fully correlated since their ratios are fixed, which explains the
thin straight-line shape of the regions as seen in the three
corresponding panels.  Compared to the results from our previous
analysis we now find that all the fluxes are clearly determined to be
non-zero, while in Refs.~\cite{Gonzalez-Garcia:2009dpj,
  Bergstrom:2016cbh} only an upper bound for the CNO fluxes was found.
This is a direct consequence of the positive evidence of neutrinos
produced in the CNO cycle provided by Borexino Phase-III spectral
data, which is here confirmed in a fully consistent global analysis.
We will discuss this point in more detail in Sec.~\ref{sec:CNO}.  We
also observe that the inclusion of the full statistics of Borexino has
improved the determination of $f_{\Nuc[7]{Be}}$ by a factor
$\mathcal{O}(3)$.

\begin{figure}\centering
  \includegraphics[width=0.95\textwidth]{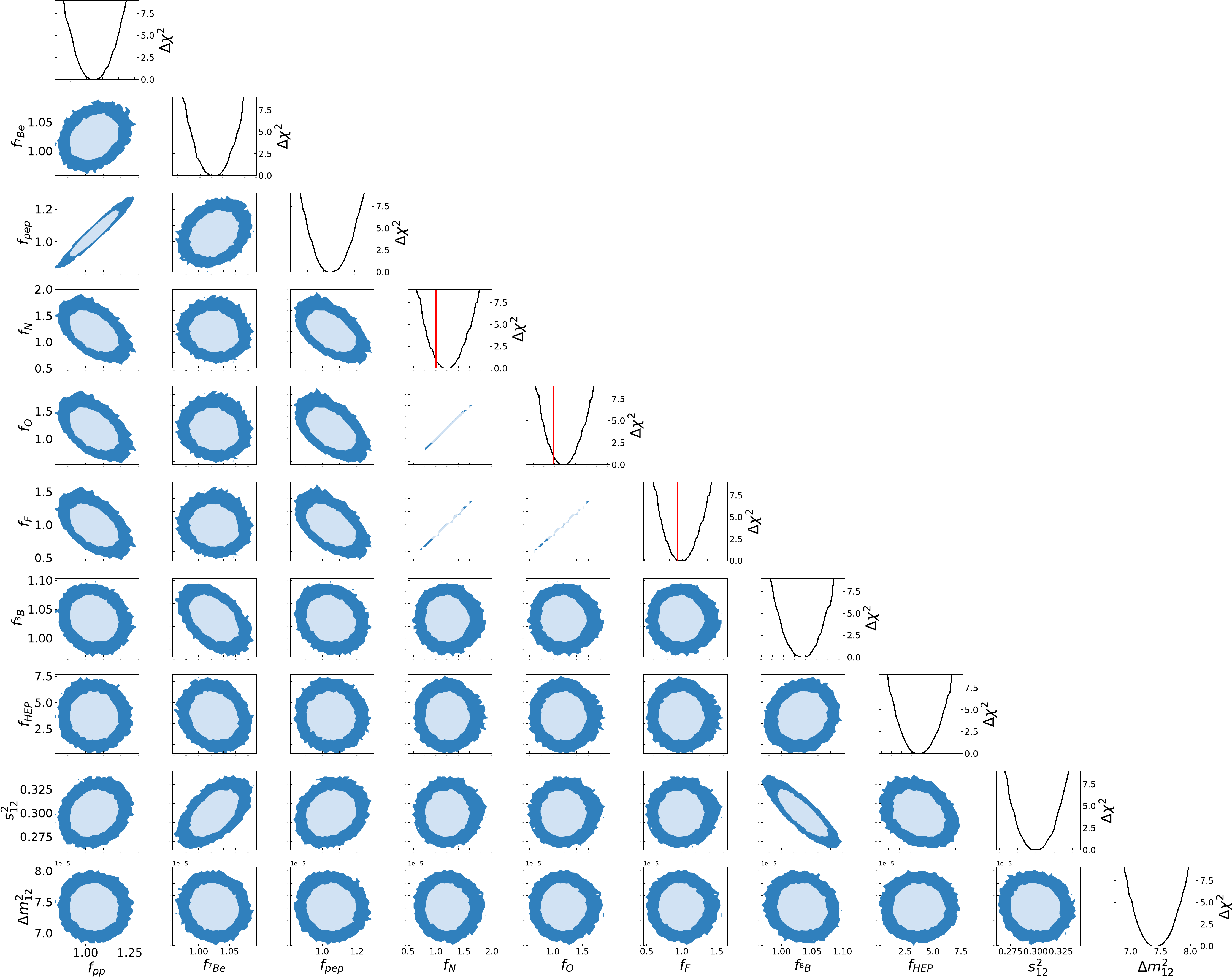}
  \caption{Same as Fig.~\ref{fig:triCNOLCGA1} but without imposing
    the luminosity constraint (see Eq.~\eqref{eq:chi2wolccno}).}
  \label{fig:triCNOwoLCGA1}
\end{figure}

Figure~\ref{fig:triCNOLCGA1} exhibits the expected correlation between
the allowed ranges of the \Nuc{pp} and \Nuc{pep} fluxes, which is a
consequence of the relation~\eqref{eq:pep-pp}.  This correlation is
somewhat weaker than what observed in the corresponding analysis in
Ref.~\cite{Bergstrom:2016cbh} because the spectral information from
Borexino Phase-II and Phase-III provides now some independent
information on $f_{\Nuc{pep}}$.  We also observe the presence of
anticorrelation between the allowed ranges of the two most intense
fluxes, \Nuc{pp} and \Nuc[7]{Be}, as dictated by the luminosity
constraint (see comparison with Fig.~\ref{fig:triCNOwoLCGA1}).
Finally we notice that the allowed ranges of $f_{\Nuc[7]{Be}}$ and
$f_{\Nuc[8]{B}}$ --~the two most precise directly determined flux
normalizations irrespective of the luminosity constraint (see
Fig.~\ref{fig:triCNOwoLCGA1})~-- are anticorrelated.  This is a direct
consequence of the different dependence of the survival probability
with $\sin^2\theta_{12}$ in their respective energy ranges.
\Nuc[8]{B} neutrinos have energies of the order of several MeV for
which the flavour transition occurs in the MSW regime and the survival
probability $P_{ee}\propto \sin^2\theta_{12}$.  Hence an increase in
$\sin^2\theta_{12}$ must be compensated by a decrease of
$f_{\Nuc[8]{B}}$ to get the correct number of events, which leads to
the anticorrelation between the $\sin^2\theta_{12}$ and
$f_{\Nuc[8]{B}}$ seen in the corresponding panel in
Fig.~\ref{fig:triCNOLCGA1}.  On the contrary, most \Nuc[7]{Be}
neutrinos have 0.86 MeV (some have 0.38 MeV) and for that energy the
flavour transition occurs in the transition regime between MSW and
vacuum average oscillations for which $P_{ee}$ decreases with
$\sin^2\theta_{12}$.  Hence the correlation between
$\sin^2\theta_{12}$ and $f_{\Nuc[7]{Be}}$ seen in the corresponding
panel.  Altogether, this leads to the anticorrelation observed between
$f_{\Nuc[7]{Be}}$ and $f_{\Nuc[8]{B}}$.  This was already mildly
present in the results in Ref.~\cite{Bergstrom:2016cbh} but it is now
a more prominent feature because of the most precise determination of
$f_{\Nuc[7]{Be}}$.

All these results imply the following share of the energy production
between the pp-chain and the CNO-cycle
\begin{equation}
  \label{eq:ppcnolum1}
  \frac{L_\text{pp-chain}}{L_\odot} =
  0.9919_{-0.0030}^{+0.0035}\, [{}_{-0.0077}^{+0.0082}]
  \quad\Longleftrightarrow\quad
  \frac{L_\text{CNO}}{L_\odot} =
  0.0079_{-0.0011}^{+0.0009}\, [{}_{-0.0026}^{+0.0028}]
\end{equation}
in perfect agreement with the SSMs which predict $L_\text{CNO} \big/
L_\odot \leq 1\%$ at the $3\sigma$ level.  Once again we notice that
in the present analysis the evidence for $L_\text{CNO}\neq 0$
clearly stands well above 99\% CL.

We next show in Fig.~\ref{fig:triCNOwoLCGA1} the results of the
analysis performed without imposing the luminosity constraint --~but
still with the ratios of the CNO fluxes fixed by the relations in
Eq.~\eqref{eq:CNOfix}~-- for which we employ
\begin{equation}
  \label{eq:chi2wolccno}
  \chi^2_\text{woLC,CNO-Rfixed} \equiv
  \chi^2_\text{global}(\vec\omega_\text{osc},\, \vec\omega_\text{flux}^\text{CNO-Rfixed})
  + \chi^2_\text{pep/pp}(f_{\Nuc{pp}},\, f_{\Nuc{pep}}) \,.
\end{equation}
The allowed ranges for the fluxes in this case are:
\begin{equation}
  \label{eq:bestwolc}
  \begin{aligned}
    f_{\Nuc{pp}}
    & = 1.038_{-0.066}^{+0.076}\, [{}_{-0.16}^{+0.18}]
    \,, \qquad
    & \Phi_{\Nuc{pp}}
    & = 6.19_{-0.39}^{+0.45}\, [{}_{-1.0}^{+1.1}]
    \times 10^{10}~\text{cm}^{-2}~\text{s}^{-1} \,,
    \\
    f_{\Nuc[7]{Be}}
    & = 1.022_{-0.018}^{+0.022}\, [{}_{-0.042}^{+0.051}]
    \,, \qquad
    & \Phi_{\Nuc[7]{Be}}
    & = 4.95_{-0.089}^{+0.11}\, [{}_{-0.22}^{+0.25}]
    \times 10^{9}~\text{cm}^{-2}~\text{s}^{-1} \,,
    \\
    f_{\Nuc{pep}}
    & = 1.039_{-0.065}^{+0.082}\, [{}_{-0.16}^{+0.19}]
    \,, \qquad
    & \Phi_{\Nuc{pep}}
    & = 1.48_{-0.09}^{+0.11}\, [{}_{-0.22}^{+0.26}]
    \times 10^{8}~\text{cm}^{-2}~\text{s}^{-1} \,,
    \\
    f_{\Nuc[13]{N}}
    & = 1.16_{-0.19}^{+0.19}\, [{}_{-0.45}^{+0.50}]
    \,, \qquad
    & \Phi_{\Nuc[13]{N}}
    & = 3.32_{-0.54}^{+0.53}\, [{}_{-1.24}^{+1.40}]
    \times 10^{8}~\text{cm}^{-2}~\text{s}^{-1} \,,
    \\
    f_{\Nuc[15]{O}}
    & = 1.16_{-0.19}^{+0.19}\, [{}_{-0.44}^{+0.49}]
    \qquad
    & \Phi_{\Nuc[15]{O}}
    & = 2.41_{-0.39}^{+0.38}\, [{}_{-0.90}^{+1.02}]
    \times 10^{8}~\text{cm}^{-2}~\text{s}^{-1} \,,
    \\
    f_{\Nuc[17]{F}}
    & = 1.01_{-0.16}^{+0.16}\, [{}_{-0.38}^{+0.45}]
    \,, \qquad
    & \Phi_{\Nuc[17]{F}}
    & = 5.25_{-0.85}^{+0.84}\, [{}_{-1.97}^{+2.21}]
    \times 10^{6}~\text{cm}^{-2}~\text{s}^{-1} \,,
    \\
    f_{\Nuc[8]{B}}
    & = 1.034_{-0.021}^{+0.020}\, [{}_{-0.051}^{+0.052}]
    \,, \qquad
    & \Phi_{\Nuc[8]{B}}
    & = 5.192_{-0.11}^{+0.10}\, [{}_{-0.26}^{+0.26}]
    \times 10^{6}~\text{cm}^{-2}~\text{s}^{-1} \,,
    \\
    f_{\Nuc{hep}}
    & = 3.6_{-1.1}^{+1.2}\, [{}_{-2.6}^{+3.0}]
    \,, \qquad
    & \Phi_{\Nuc{hep}}
    & = 2.9_{-0.9}^{+1.0}\, [{}_{-2.1}^{+2.4}]
    \times 10^{4}~\text{cm}^{-2}~\text{s}^{-1} \,.
  \end{aligned}
\end{equation}
As expected, the \Nuc{pp} flux is the most affected by the release of
the luminosity constraint as it is this reaction which gives the
largest contribution to the solar energy production and therefore its
associated neutrino flux is the one more strongly bounded when
imposing the luminosity constraint.  The \Nuc{pep} flux is also
affected due to its strong correlation with the \Nuc{pp} flux,
Eq.~\eqref{eq:pep-pp}.  The CNO fluxes are mildly affected in an
indirect way due to the modified contribution of the \Nuc{pep} fluxes
to the Borexino spectra.

Thus we find that the energy production in the pp-chain and the
CNO-cycle without imposing the luminosity constraint are given by:
\begin{equation}
  \label{eq:ppcnolum2}
  \frac{L_\text{pp-chain}}{L_\odot}
  = 1.030_{-0.061}^{+0.070}\, [{}_{-0.15}^{+0.17}]
  \qquad\text{and}\qquad
  \frac{L_\text{CNO}}{L_\odot}
  = 0.0075_{-0.0013}^{+0.0013}\, [{}_{-0.0029}^{+0.0030}] \,.
\end{equation}
Comparing Eqs.~\eqref{eq:ppcnolum1} and~\eqref{eq:ppcnolum2} we see
that while the amount of energy produced in the CNO cycle is about the
same in both analysis, releasing the luminosity constraint allows for
larger production of energy in the pp-chain.  So in this case we find
that the present value of the ratio of the neutrino-inferred solar
luminosity, $L_\odot\text{(neutrino-inferred)}$, to the photon
measured luminosity $L_\odot$ is:
\begin{equation}
  \label{eq:lnutot}
  \frac{L_\odot\text{(neutrino-inferred)}}{L_\odot}
  = 1.038_{-0.060}^{+0.069}\, [{}_{-0.15}^{+0.17}] \,.
\end{equation}
The neutrino-inferred luminosity is in good agreement with the one
measured in photons, with a $1\sigma$ uncertainty of $\sim 6\%$.  This
represents only a very small variation with respect to the previous
best determination~\cite{Bergstrom:2016cbh}.  Such result is expected
because the determination of the \Nuc{pp} flux, which, as mentioned
above gives the largest contribution to the neutrino-inferred solar
luminosity, has not improved sensibly with the inclusion of the full
statistics of the phases II and III of Borexino.

\begin{figure}\centering
  \includegraphics[width=0.95\textwidth]{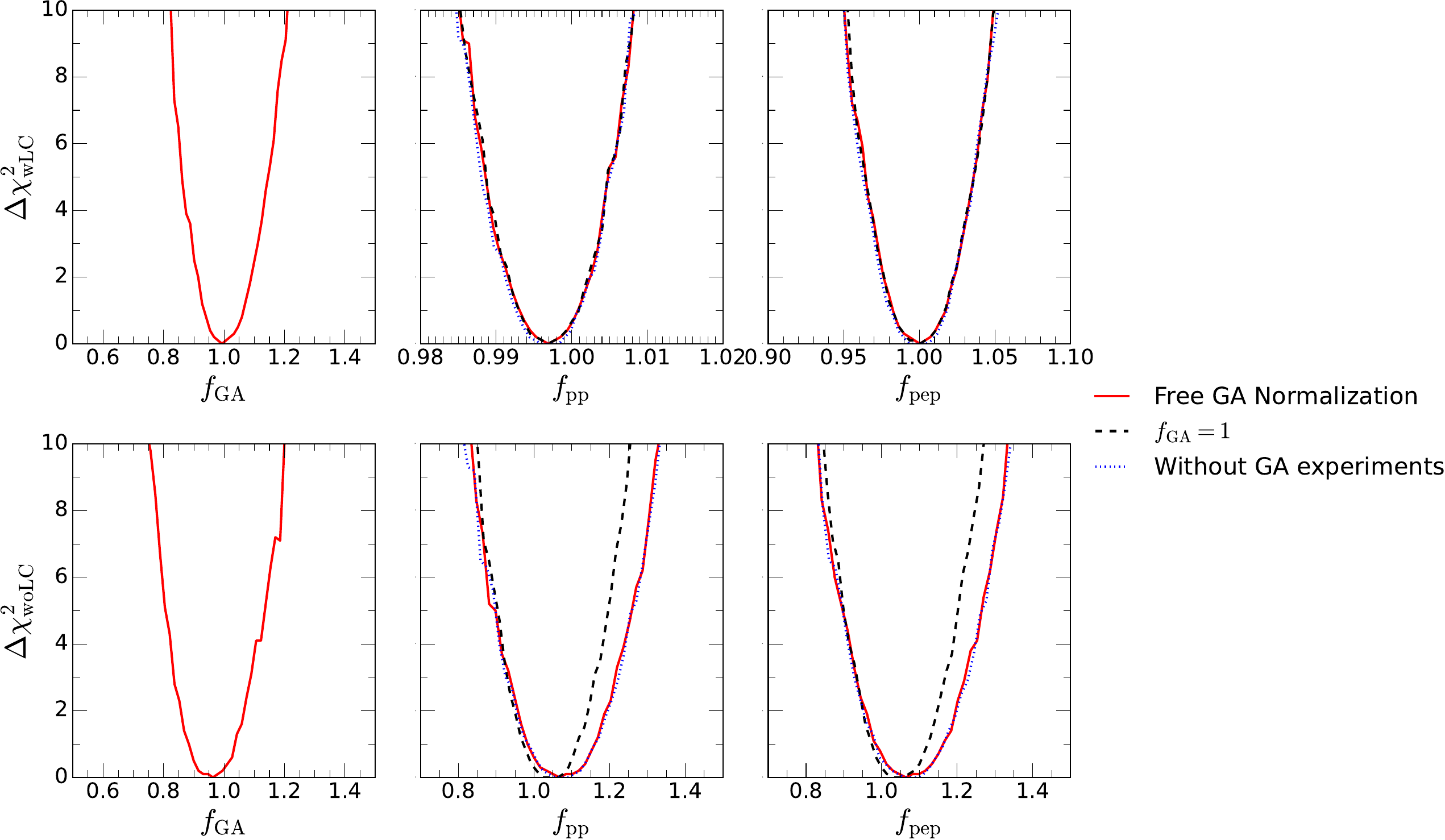}
  \caption{Dependence of the determination of the \Nuc{pp} and \Nuc{pep}
    fluxes on the assumptions about the Gallium experiments included
    in the analysis.  The upper (lower) panels show
    the results of different variants of the analysis <<CNO-Rfixed>>
    with (without) luminosity constraint.  See text for details.}
  \label{fig:gallium}
\end{figure}

We finish this section by discussing the role of the Gallium
experiments in these results with the aim of addressing the possible
impact of the Gallium anomaly~\cite{Laveder:2007zz, Acero:2007su,
  Giunti:2010zu}.  As described in the introduction, this anomaly
consists in a deficit of the event rate observed in Gallium source
experiments with respect to the expectation, which represents an
obvious puzzle for the interpretation of the results of the solar
neutrino Gallium measurements.  In this work we assume the well
established standard $3\nu$ oscillation scenario and in this context
the attempts at explanation (or at least alleviation) of the anomaly
invoke the uncertainties of the capture cross
section~\cite{Berryman:2021yan, Giunti:2022xat, Elliott:2023xkb}.
Thus the open question posed by the Gallium anomaly is the possible
impact of such modification of the cross section in the results of our
fit.

In order to quantify this we performed two additional variants of our
analysis.  In the first one we introduce an additional parameter,
$f_\text{GA}$, which multiplies the predicted event rates from all
solar fluxes in the Gallium experiments.  This parameter is left free
to vary in the fits and would mimic an energy independent modification
of the capture cross section (or equivalently of the detection
efficiency).  In the second variant we simply drop Gallium experiments
from our global fit.

The results of these explorations are shown in Fig.~\ref{fig:gallium}
where we plot the most relevant marginalized one-dimensional
projections of $\Delta\chi^2$ for these two variants.  The upper
(lower) panels correspond to analysis performed with (without) the
luminosity constraint.  The left panel shows the projection over the
normalization parameter $f_\text{GA}$ obtained in the variant of the
analysis which makes use of this parameter.  As seen from the figure,
the results of the fit favour $f_\text{GA}$ close to one, or, in other
words, the global analysis of the solar experiments do \emph{not}
support a modification of the neutrino capture cross section in
Gallium (or any other effect inducing an energy-independent reduction
of the detection efficiency in the Gallium experiments).  This is so
because, within the $3\nu$ oscillation scenario, the global fit
implies a rate of \Nuc{pp} and \Nuc[7]{Be} neutrinos in the Gallium
experiment which is in good agreement with the luminosity constraint
as well as with the rates observed in Borexino.

On the central and right panel of the figure we show the corresponding
modification of marginalized one-dimensional projections of
$\Delta\chi^2$ on the \Nuc{pp} and \Nuc{pep} flux normalizations which
are those mostly affected in these variants.  For the sake of
comparison, in the upper and lower panels we also plot the results for
the $f_\text{GA} = 1$ analysis (also visible in the corresponding
windows in Figs.~\ref{fig:triCNOLCGA1} and~\ref{fig:triCNOwoLCGA1},
respectively).  The figure illustrates that once the luminosity
constraint is imposed, the determination of the solar fluxes is
totally unaffected by the assumptions about the capture rate in
Gallium.  As seen in the lower panels, even without the luminosity
constraint the impact on the \Nuc{pp} and \Nuc{pep} determination is
marginal, which emphasizes the robustness of the flux determination in
Eqs.~\eqref{eq:bestlc} and~\eqref{eq:bestwolc}.  This is the case
thanks to the independent precise determination of the \Nuc{pp} flux
in the phases I and II in Borexino.  Furthermore, the small
modification is the same irrespective of whether the Gallium capture
rate is left free or completely removed from the analysis; this is due
to the lack of spectral and day-night capabilities in Gallium
experiments, which prevents them from providing further information
beyond the overall normalization scale of the signal.

\section{Examination of the determination of the CNO fluxes}
\label{sec:CNO}

As mentioned above, one of the most important developments in the
experimental determination of the solar neutrino fluxes in the last
years have been the evidence of neutrinos produced in the CNO cycle
reported by Borexino~\cite{BOREXINO:2020aww, BOREXINO:2022abl,
  Borexino:2023puw}.  The detection was made possible thanks to a
novel method to constrain the rate of the \Nuc[210]{Bi} background.
In Ref.~\cite{BOREXINO:2020aww}, using a partial sample of their
Phase-III data, the collaboration found a $5.1\sigma$ significance of
the CNO flux observation, which increased to $7\sigma$ with the full
Phase-III statistics~\cite{BOREXINO:2022abl}, and to about $8\sigma$
when combined with the CID method~\cite{Borexino:2023puw}.  See
appendix~\ref{app:borex} for details.

Key ingredients in the analysis performed by the collaboration in
Refs.~\cite{BOREXINO:2020aww, BOREXINO:2022abl, Borexino:2023puw} (and
therefore in the derivation of these results) are the assumptions
about the relative contribution of the three reactions producing
neutrinos in the CNO cycle, as well as those about other solar fluxes
in the same energy range, in particular the \Nuc{pep} neutrinos.  In a
nutshell, as mentioned above, the collaboration assumes a common shift
of the normalization of the CNO fluxes with respect to that of the
SSM, and it is the evidence of a non-zero value of such normalization
which is quantified in Refs.~\cite{BOREXINO:2020aww, BOREXINO:2022abl,
  Borexino:2023puw}.  In what respects the rate from the \Nuc{pep}
flux, the SSM expectation was assumed because the Phase-III data by
itself does not allow to constraint simultaneously the CNO and
\Nuc{pep} flux normalizations.

\begin{figure}\centering
  \includegraphics[width=0.95\textwidth]{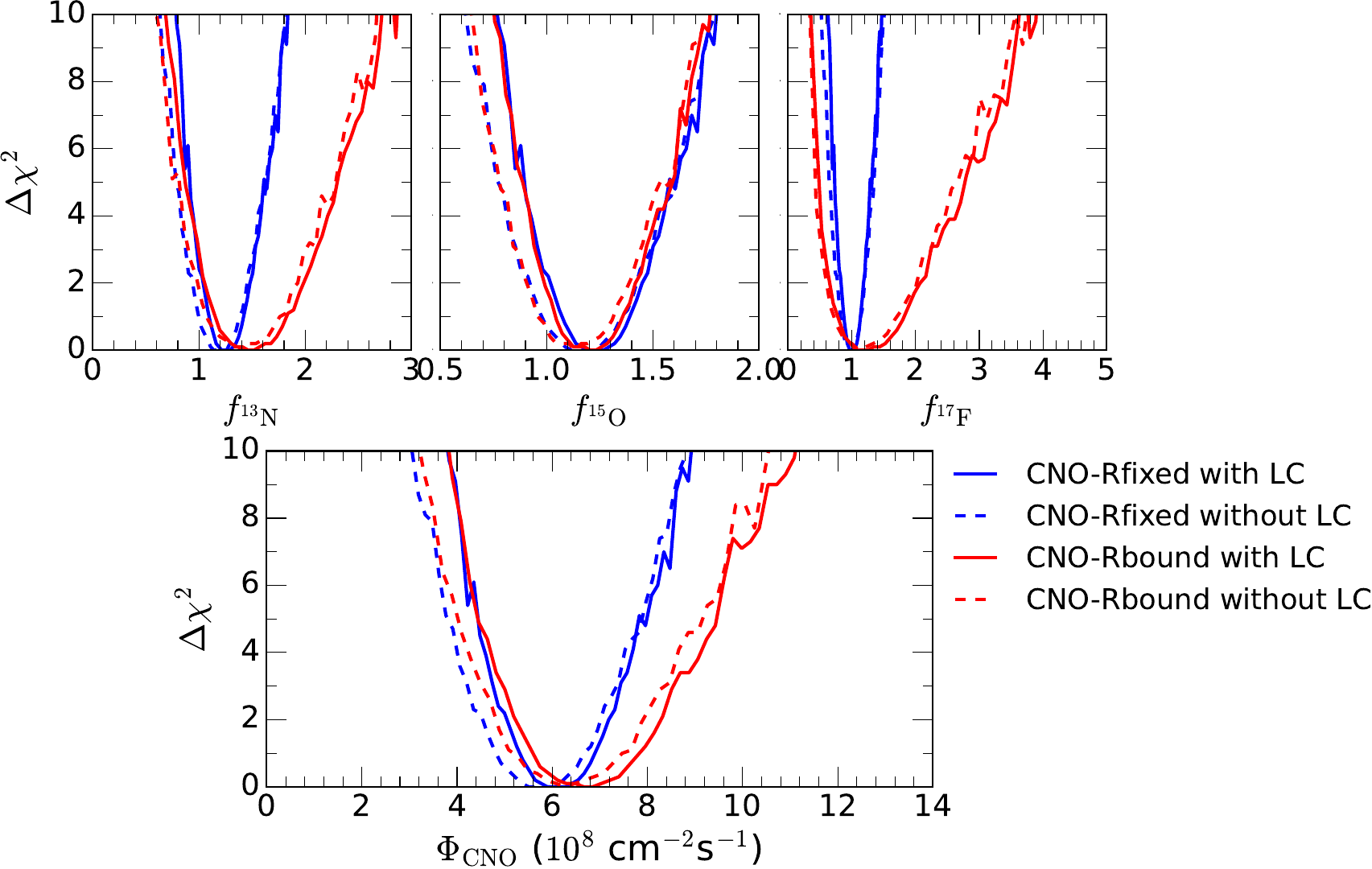}
  \caption{One dimensional projections of the global $\Delta\chi^2$ of
    the for the three neutrino fluxes produced in the CNO-cycle for
    different assumptions as labeled in the figure.  See text for
    details.}
  \label{fig:CNO}
\end{figure}

In this respect, the global analysis presented in the previous section
are performed under the same paradigm of a common shift normalization
of the CNO fluxes, but being global, the \Nuc{pep} flux normalization
is also simultaneously fitted.  For the sake of comparison we
reproduce in Fig.~\ref{fig:CNO} the projection of the marginalized
$\Delta\chi^2_\text{wLC,CNO-Rfixed}$~\eqref{eq:chi2wlccno} and
$\Delta\chi^2_\text{woLC,CNO-Rfixed}$~\eqref{eq:chi2wlccno} on the
normalization parameters for the three CNO fluxes.  For convenience we
also show the projections as a function of the total neutrino flux
produced in the CNO cycle.  As seen in the figure the results of the
analysis (either with or without luminosity constraint) yield a value
of $\Delta\chi^2$ well beyond $3\sigma$ for $\Phi_{\Nuc{CNO}}=0$.  A
dedicated run for this no-CNO scenario case gives $\Delta\chi^2 = 54$
($33$) for the analysis with (without) luminosity constraints, and it
is therefore excluded at $7.3\sigma$ ($5.7\sigma$) CL.

In order to study the dependence of the results on the assumption of a
unique common shift of the normalization of three CNO fluxes we
explored the possibility of making a global analysis in which the
three normalization parameters are varied independently.  As mentioned
above, a priori the three normalizations only have to be subject to a
minimum set of consistency relations in Eqs.~\eqref{eq:CNOineq1}
and~\eqref{eq:CNOineq2}.  However, as discussed in detail in
Sec.~\ref{sec:bx3nfit}, the background model in
Refs.~\cite{BOREXINO:2020aww, BOREXINO:2022abl, Borexino:2023puw} only
assumes an upper bound on the amount of \Nuc[210]{Bi} and cannot be
reliably employed to such general analysis because of the larger
degeneracy between the \Nuc[210]{Bi} background and the \Nuc[13]{N}
flux spectra.

With this limitation in mind, we proceed to perform two alternative
analysis (with and without imposing the luminosity constraints) in
which the normalization of the three CNO fluxes are left free to vary
independently but with ratios constrained within a range broad enough
to generously account for all variants of the B23 SSM, but not to
extend into regions of the parameter space where the assumptions on
the background model may not be applicable.
Conservatively neglecting correlations between their theoretical
uncertainties, the neutrino fluxes of SSMs presented in
Ref.~\cite{Magg:2022rxb} and available publicly through a public
repository \cite{B23Fluxes} verify
\begin{equation}
  \label{eq:cnoRanges}
  \frac{f_{\Nuc[15]{O}}}{f_{\Nuc[13]{N}}} =
  \begin{cases}
    1.00\,(1\pm 0.24) \\
    0.95\,(1\pm 0.22) \\
    0.96\,(1\pm 0.21) \\
    1.01\,(1\pm 0.23) \\
    1.00\,(1\pm 0.23)
  \end{cases}
  \qquad
  \frac{f_{\Nuc[17]{F}}}{f_{\Nuc[13]{N}}} =
  \begin{cases}
    1.00\,(1\pm 0.25) &\text{B23-GS98} \\
    0.84\,(1\pm 0.23) &\text{B23-AGSS09-met} \\
    0.80\,(1\pm 0.20) &\text{B23-AAG21} \\
    0.79\,(1\pm 0.22) &\text{B23-MB22-met} \\
    0.79\,(1\pm 0.22) &\text{B23-MB22-phot}
  \end{cases}
\end{equation}
Thus in these analyses, here onward labeled <<CNO-Rbound>>, we
introduce two pulls $\xi_1$ and $\xi_2$ for these two ratios.  Notice,
however, that we could have equally defined the priors with respect to
the reciprocal of the ratios in Eq.~\eqref{eq:cnoRanges}.  Hence, in
order to avoid a bias towards larger fluxes in the numerator versus
the denominator introduced by either choice, we resort instead to
logarithmic priors for the ratios:
\begin{multline}
  \vec\omega_\text{flux}^\text{CNO-Rbound} \equiv
  (f_{\Nuc{pp}},\; f_{\Nuc[7]{Be}},\; f_{\Nuc{pep}},\; f_{\Nuc[13]{N}},
  \\
  f_{\Nuc[15]{O}} = 0.98\, \exp(\xi_1) \, f_{\Nuc[13]{N}},\;
  f_{\Nuc[17]{F}} = 0.85\, \exp(\xi_2)\, f_{\Nuc[13]{N}},\;
  f_{\Nuc[8]{B}},\; f_{\Nuc{hep}})
\end{multline}
and add two Gaussian penalty factors for these pulls, so that the
corresponding $\chi^2$ function without the luminosity constraint is:
\begin{equation}
  \chi^2_\text{woLC,CNO-Rbound}
  \equiv \chi^2_\text{global}(\vec\omega_\text{osc},\, \vec\omega_\text{flux}^\text{CNO-Rbound})
  + \chi^2_\text{pep/pp}(f_{\Nuc{pp}},\, f_{\Nuc{pep}})
  + \frac{\xi_1^2}{\sigma_{\xi_1}^2} + \frac{\xi_2^2}{\sigma_{\xi_2}^2}
\end{equation}
with $\sigma_{\xi_1}=0.26$ and $\sigma_{\xi_2}=0.48$, chosen to cover
the ranges in Eq.~\eqref{eq:cnoRanges}.  In addition
$f_{\Nuc[13]{N}}$, $f_{\Nuc[15]{O}}$, and $f_{\Nuc[17]{F}}$ are
required to verify the consistency relations in
Eqs.~\eqref{eq:CNOineq1} and~\eqref{eq:CNOineq2}.  The $\chi^2$
function with the luminosity constraint is obtained by further
including the $\chi^2_\text{LC}$ prior of Eq.~\eqref{eq:priorLC}:
\begin{equation}
  \label{eq:chi2wlcnmod}
  \chi^2_\text{wLC,CNO-Rbound}
  \equiv \chi^2_\text{woLC,CNO-Rbound}
  + \chi^2_\text{LC}(\vec\omega_\text{flux}^\text{CNO-Rbound}) \,.
\end{equation}

We plot in Fig.~\ref{fig:CNO} the projection of the marginalized
$\Delta\chi^2_\text{wLC,CNO-Rbound}$~\eqref{eq:chi2wlcnmod} and
$\Delta\chi^2_\text{woLC,CNO-Rbound}$~\eqref{eq:chi2wlcnmod} on the
normalization parameters for the three CNO fluxes as well as on the
total neutrino flux produced in the CNO cycle.  As seen in the figure,
allowing for the ratios of the CNO normalizations to vary within the
intervals~\eqref{eq:cnoRanges} has little impact on the allowed range
of the \Nuc[15]{O} flux and on the lower limit of the \Nuc[13]{N} and
\Nuc[17]{F} fluxes.  As a consequence, the CL at which the no-CNO
scenario can be ruled out is unaffected.  On the contrary, we see in
Fig.~\ref{fig:CNO} that the upper bound on the \Nuc[13]{N} and
\Nuc[17]{F} fluxes, and therefore of the total neutrino flux produced
in the CNO-cycle, is relaxed.\footnote{The allowed ranges for the
fluxes produced in the pp-chain are not substantially modified with
respect to the ones obtained from the <<CNO-Rfixed>> fits,
Eqs.~\eqref{eq:bestlc}~and~\eqref{eq:bestwolc}.}  This is a
consequence of the strong degeneracy between the spectrum of events
from these fluxes and those from the \Nuc[210]{Bi} background
mentioned above, see Fig.~\ref{fig:subspeccomp} and discussion in
Sec.~\ref{sec:bx3nfit}.  Conversely the fact that the range of the
\Nuc[15]{O} flux is robust under the relaxation of the constraints on
the CNO flux ratios, means that the high statistics spectral data of
the Phase-III of Borexino holds the potential to differentiate the
event rates from \Nuc[15]{O} $\nu$'s from those from \Nuc[13]{N} and
\Nuc[17]{F} $\nu$'s.  The reliable quantification of this possibility,
however, requires the knowledge of the minimum allowed value of the
\Nuc[210]{Bi} background which so far has not been presented by the
collaboration.

So, let us emphasize that our <<CNO-Rbound>> analysis have been
performed with the aim of testing the effect of relaxing the severe
constrains on the CNO fluxes in the studies of the Borexino
collaboration.  Our conclusion is that the statistical significance of
the evidence of detection of events produced by neutrinos from the
CNO-cycle is affected very little by the relaxation of the constraint
on their relative ratios.  However, their allowed range is, and this
can have an impact when confronting the results of the fit with the
predictions of the SSM as we discuss next.

\section{Comparison with Standard Solar Models}
\label{sec:compaSSM}

Next we compare the results of our determination of the solar fluxes
with the expectations from the five B23 solar models: SSMs computed
with the abundances compiled in table 5 of~\cite{Magg:2022rxb} based
on the photospheric and meteoritic solar mixtures (MB22-phot and
MB22-met models, respectively), and with the~\cite{Asplund2021}
(AAG21), the meteoritic scale from~\cite{Asplund2009} (AGSS09-met),
and~\cite{Grevesse1998} (GS98) compositions.  We use both MB22-phot
and MB22-met for completeness, although the abundances are very
similar in both scales, as clearly reflected by the results in this
section.  A similar agreement would be found using both the meteoritic
and photospheric scales from AAG21, and therefore we use only one
scale in this case.\footnote{The structures of these models, as well
as the total neutrino fluxes and internal distributions are available
at \cite{B23Fluxes}.}

\begin{figure}\centering
  \includegraphics[width=0.95\textwidth]{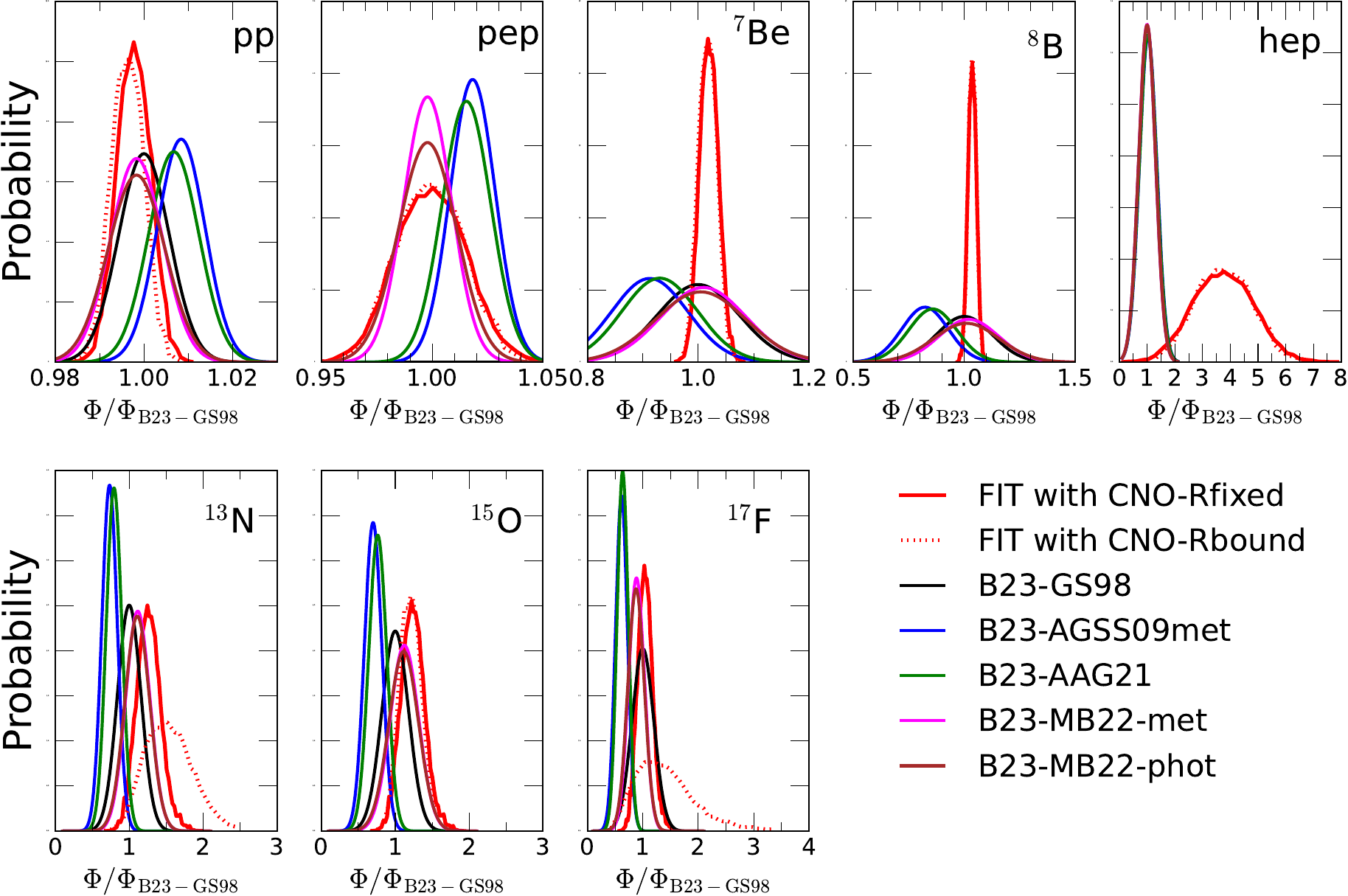}
  \caption{Marginalized one-dimensional probability distributions for
    the best determined solar fluxes in our analysis as compared to
    the predictions for the five SSMs in Ref.~\cite{Magg:2022rxb,B23Fluxes}.}
  \label{fig:compaSSM}
\end{figure}

SSMs predict that nuclear energy accounts for all the solar luminosity
(barred about a few parts in $10^4$ that are of gravothermal origin)
so for all practical matters the neutrino fluxes predicted by SSMs
satisfy the luminosity constraint.  Therefore we compare the
expectations of the various SSM models with the results of our
analysis performed with such constraint.  In what respects the
assumptions on the CNO fluxes, in order to explore the dependence of
our conclusions on the specific choice of flux ratios we quantify the
results obtained in both the <<FIT=CNO-Rfixed>> analysis (with
$\chi^2_\text{FIT}$ in Eq.~\eqref{eq:chi2wlccno}) and the
<<FIT=CNO-Rbounded>> one (with $\chi^2_\text{FIT}$ in
Eq.~\eqref{eq:chi2wlcnmod}).
For illustration we plot in Fig.~\ref{fig:compaSSM} the marginalized
one-dimensional probability distributions for the best determined
solar fluxes in such two cases as compared to the predictions for the
five B23 SSMs.  The probability distributions for our fits are
obtained from the one-dimensional marginalized
$\Delta\chi^2_\text{FIT}(f_i)$ of the corresponding analysis as
$P_\text{FIT}(f_i)\propto \exp[-\Delta\chi^2_\text{FIT}(f_i)/2]$
normalized to unity.  To construct the analogous distributions for
each of the SSMs we use the predictions $\langle f_i^\text{SSM}
\rangle$ for the fluxes, the relative uncertainties
$\sigma_i^\text{SSM}$ and their correlations $\rho_{ij}^\text{SSM}$ as
obtained from Refs.~\cite{B23Fluxes}, and also assume gaussianity so
to build the corresponding
$\chi^2_{\text{SSM}}(\vec\omega_\text{flux})$
\begin{equation}
  \label{eq:chi2mod}
  \chi^2_{\text{SSM}}(\vec\omega_\text{flux})=\sum_{i,j}
  (f_i-f_i^\text{SSM}) C^{-1}_{ij} (f_i-f_i^\text{SSM})
  \quad\text{with}\quad
  C_{ab} = \sigma_a^\text{SSM} \sigma_b^\text{SSM}\rho_{ab} \,,
\end{equation}
from which it is trivial to obtain the marginalized one-dimensional
$\Delta\chi^2_\text{SSM}(f_i)$ and construct the probability
$P_\text{SSM}(f_i)\propto \exp[-\Delta\chi^2_\text{SSM}(f_i)/2]$.

In the frequentist statistical approach, quantitative comparison of a
model prediction for a set of fluxes with the results from the data
analysis can be obtained using the \emph{parameter goodness of fit}
(PG) criterion introduced in Ref.~\cite{Maltoni:2003cu}, by comparing
the minimum value of $\chi^2$ function for the analysis of the data
with that obtained for the same analysis adding the prior imposed by
the model.\footnote{In this respect it is important to notice that, in
order to avoid any bias towards one of the models in the data
analysis, in both <<CNO-Rfixed>> and <<CNO-Rbound>> cases the
assumptions on the ratios of the three CNO fluxes have been chosen to
be ``model-democratic'', \textit{i.e.}, centered at the average of the
predictions of the models (and, in the case of <<CNO-Rbound>>, with
$1\sigma$ uncertainties covering the $1\sigma$ range allowed by all
SSM models).}  Thus, following Ref.~\cite{Maltoni:2003cu}, we
construct the test statistics
\begin{multline}
  \label{eq:dchifitmod}
  \Delta\chi^2_\text{FIT,SSM,SET}
  = \left.
  \big[\chi^2_\text{FIT}(\vec\omega_\text{osc},\, \vec\omega_\text{flux}^\text{FIT})
    + \chi^2_\text{SSM,SET}(\vec\omega_\text{flux}^\text{FIT}) \big]
  \right|_\text{min}
  \\
  - \left. \chi^2_\text{FIT}(\vec\omega_\text{osc},\, \vec\omega_\text{flux}^\text{FIT})
  \right|_\text{min}
  - \left. \chi^2_\text{SSM,SET}(\vec\omega_\text{flux}^\text{FIT})
  \right|_\text{min}
\end{multline}
where $\chi^2_\text{SSM,SET}(\vec\omega_\text{flux})$ is obtained as
Eq.~\eqref{eq:chi2mod} with $i,j$ (and $a,b$) fluxes restricted to a
specific subset as specified by ``SET''.  The minimization of each of
the terms in Eq.~\eqref{eq:dchifitmod} is performed independently in
the corresponding parameter space.  $\Delta\chi^2_\text{FIT,SSM,SET}$
follows a $\chi^2$ distribution with $n$ degrees of freedom, which, in
the present case, coincides with the number of free parameters in
common between $\chi^2_\text{FIT}(\vec\omega_\text{osc},\,
\vec\omega_\text{flux}^\text{FIT})$ and
$\chi^2_\text{SSM,SET}(\vec\omega_\text{flux}^\text{FIT})$.  Notice
that, by construction, the result of the test depends on the number of
fluxes to be compared, \textit{i.e.}, on the fluxes in ``SET'', both
because of the actual comparison between the measured and predicted
values for those specific fluxes, and because of the change in $n$
with which the $p$-value of the model is to be computed.  This is
illustrated in Table~\ref{tab:pgf} where we list the values of
$\Delta\chi^2_\text{FIT,SSM,SET}$ for different choices of ``SET''
which we have labeled as:
\begin{equation}
  \begin{tabular}{ll}
    SET  & constrained fluxes
    \\
    \hline
    FULL
    & ($f_{\Nuc{pp}}$, $f_{\Nuc[7]{Be}}$,\, $f_{\Nuc{pep}}$, $f_{\Nuc[13]{N}}$,
    $f_{\Nuc[15]{O}}$, $f_{\Nuc[17]{F}}$, $f_{\Nuc[8]{B}}$, $f_{\Nuc{hep}}$)
    \\
    Be+B+CNO
    & ($f_{\Nuc[7]{Be}}$, $f_{\Nuc[13]{N}}$, $f_{\Nuc[15]{O}}$,
    $f_{\Nuc[17]{F}}$, $f_{\Nuc[8]{B}}$)
    \\
    CNO & ($f_{\Nuc[13]{N}}$, $f_{\Nuc[15]{O}}$, $f_{\Nuc[17]{F}}$)
  \end{tabular}
\end{equation}

\begin{table}\centering
  \catcode`?=\active\def?{\hphantom{0}}
  \begin{footnotesize}
    \begin{tabular}{|c|l|ccc|ccc|ccc|}
      \hline
      FIT & B23-SSM & \multicolumn{3}{c|}{FULL}
      & \multicolumn{3}{c|}{Be+B+CNO} & \multicolumn{3}{c|}{CNO}
      \\
      \hline
      \multirow{6}{*}{\rotatebox{90}{CNO-Rfixed}}
      &
      & \multicolumn{3}{c|}{n=6}
      & \multicolumn{3}{c|}{n=3}
      & \multicolumn{3}{c|}{n=1}
      \\
      \cline{3-11}
      &
      & $\Delta\chi^2$ & $p_\text{GF}$ & CL [$\sigma$]
      & $\Delta\chi^2$ & $p_\text{GF}$ & CL [$\sigma$]
      & $\Delta\chi^2$ & $p_\text{GF}$ & CL [$\sigma$]
      \\
      \cline{2-11}
      & AGSS09-met    & 14.5 & 0.024 & 2.3 & 9.8 & 0.020 & 2.3? & 7.2 & 0.0073 & 2.7 \\
      & GS98          & ?8.1 & 0.24? & 1.2 & 3.0 & 0.39? & 0.86 & 2.4 & 0.12?? & 1.5 \\
      & AAG21         & 12.5 & 0.052 & 1.9 & 7.8 & 0.05? & 2.0? & 6.2 & 0.013? & 2.5 \\
      & MB22-met/phot & ?7.1 & 0.31? & 1.0 & 2.2 & 0.53? & 0.62 & 2.0 & 0.16?? & 1.4
      \\
      \hline
      \multirow{6}{*}{\rotatebox{90}{CNO-Rbound}}
      &
      & \multicolumn{3}{c|}{n=8}
      & \multicolumn{3}{c|}{n=5}
      & \multicolumn{3}{c|}{n=3}
      \\
      \cline{3-11}
      &
      & $\Delta\chi^2$ & $p_\text{GF}$ & CL [$\sigma$]
      & $\Delta\chi^2$ & $p_\text{GF}$ & CL [$\sigma$]
      & $\Delta\chi^2$ & $p_\text{GF}$ & CL [$\sigma$]
      \\
      \cline{2-11}
      & AGSS09-met    & 14.1 & 0.079 & 1.8? & 9.3 & 0.098 & 1.7? & 7.2 & 0.066 & 1.8? \\
      & GS98          & ?6.7 & 0.57? & 0.57 & 1.7 & 0.88? & 0.14 & 1.6 & 0.66? & 0.44 \\
      & AAG21         & 11.7 & 0.16? & 1.4? & 6.8 & 0.24? & 1.2? & 5.7 & 0.13? & 1.5? \\
      & MB22-met/phot & ?5.9 & 0.66? & 0.44 & 1.1 & 0.95? & 0.06 & 1.0 & 0.80? & 0.25 \\
      \hline
    \end{tabular}
  \end{footnotesize}
  \caption{Results of the PG test for the different models and data
    samples considered.  Within the given accuracy the results for
    MB22-met and MB22-phot models are the same.}
  \label{tab:pgf}
\end{table}

Upon analyzing the data in the Table~\ref{tab:pgf}, it becomes evident
that the B23-MB22 models (both the meteoritic and photospheric
variations) exhibit a significantly higher level of compatibility with
the observed data, even slightly better the B23-GS98 model.  On the
contrary the B23-AGSS09met and B23-AAG21 models exhibits a lower level
of compatibility with observations, with B23-AAG21 model slightly
better aligned with the data.
Maximum discrimination is provided by comparing mainly the CNO fluxes
for which the prediction of both models is mostly different.  On the
other hand, including all the fluxes from the pp-chain in the
comparison tends to dilute the discriminating power of the test.  The
table also illustrates how allowing for the three CNO fluxes
normalizations to vary in the fit tends to relax the CL at which the
models are compatible with the observations.

Let us remember that our previously determined fluxes in
Ref~\cite{Bergstrom:2016cbh} when confronted with the GS98 and AGSS09
models of the time~\cite{Serenelli:2011py} showed \emph{absolutely} no
preference for either model.  This was driven by the fact that the
most precisely measured \Nuc[8]{B} flux (and also of \Nuc[7]{Be}) laid
right in the middle of the prediction of both models.  The new
B16-GS98 model in Ref.~\cite{Vinyoles:2016djt} predicted a slightly
lower value for \Nuc[8]{B} flux in slightly better agreement with the
extracted fluxes of Ref~\cite{Bergstrom:2016cbh}, but the conclusion
was still that there was no significant preference for either model.
Compared to those results, both the most precisely determined
\Nuc[7]{Be} flux and, most importantly, the newly observed rate of CNO
events in Borexino have consistently moved towards the prediction of
the models with higher metallicity abundances.

Let us finish commenting on the relative weight of the experimental
precision versus the theoretical model uncertainties in the results in
Table~\ref{tab:pgf}.  To this end one can envision an ideal experiment
which measures $f_i$ to match precisely the values predicted by one of
the models with infinite accuracy.  Assuming the measurements to
coincide with the predictions of B23-GS98, one gets
$\Delta\chi^2_\text{SSM,SET} = 17.1$ and $16.7$ for SSM=B23-AGSS09-met
and SSM=B23-AAG21 with SET=FULL, which means that the maximum CL at
which these two SSM can be disfavoured is $2.2\sigma$ and $2.1\sigma$.
Choosing instead SET=CNO these numbers become
$\Delta\chi^2_\text{SSM,SET} = 15.0$ and $14.1$ for SSM=B23-AGSS09-met
and SSM=B23-AAG21, respectively, corresponding to a $3.1\sigma$ and
$3.0\sigma$ maximum rejection.
This stresses the importance of reducing the uncertainties in the
model predictions to boost the discrimination between the models.

\section{Summary}
\label{sec:summary}

In this work we have updated our former determination of solar
neutrino fluxes from neutrino data as presented in
Refs.~\cite{Gonzalez-Garcia:2009dpj, Bergstrom:2016cbh}, by
incorporating into the analysis the latest results from both solar and
non-solar neutrino experiments.  In particular this includes the full
data from the three phases of the Borexino experiments which have
provided us with the first direct evidence of neutrinos produced in
the CNO-cycle.

We have derived the best neutrino oscillation parameters and solar
fluxes constraints using a frequentist analysis with and without
imposing nuclear physics as the only source of energy generation
(luminosity constraint).  Compared to the results from previous
analysis we find that the determination of the \Nuc[7]{Be} flux has
improved by a factor $\mathcal{O}(3)$, but most importantly we now
find that the three fluxes produced in the CNO-cycle are clearly
determined to be non-zero, with $1\sigma$ precision ranges between
20\% to $\sim 100\%$ depending on the assumptions in the analysis
about their relative normalization.  Conversely, in
Refs.~\cite{Gonzalez-Garcia:2009dpj, Bergstrom:2016cbh} only an upper
bound for the CNO fluxes was found.  This also implies that it is
solidly established that at 99\% CL the solar energy produced in the
CNO-cycle is between $0.46\%$ and $1.05\%$ of the total solar
luminosity.

The observation of the CNO neutrinos is also paramount to discriminate
among the different versions of the SSMs built with different inputs
for the solar abundances, since the CNO fluxes are the most sensitive
to the solar composition.  In this work we confront for the first time
the neutrino fluxes determined on a purely experimental basis with the
predictions of the latest generation of SSM obtained in
Ref.~\cite{Magg:2022rxb,B23Fluxes}.  Our results show that the SSMs built incorporating
lower metallicities are less compatible with the solar neutrino
observations.

\acknowledgments

This project is funded by USA-NSF grant PHY-1915093 and by the
European Union through the Horizon 2020 research and innovation
program (Marie Sk{\l}odowska-Curie grant agreement 860881-HIDDeN) and
the Horizon Europe programme (Marie Sk{\l}odowska-Curie Staff Exchange
grant agreement 101086085-ASYMMETRY).  It also receives support from
grants PID2019-\allowbreak 110058GB-\allowbreak C21,
PID2019-\allowbreak 105614GB-\allowbreak C21, PID2019-\allowbreak
108892RB-\allowbreak I00, PID2019-\allowbreak 110058GB-\allowbreak
C21, PID2020-\allowbreak 113644GB-\allowbreak I00, PID2022-\allowbreak
142545NB-\allowbreak C21, ``Unit of Excellence Maria de Maeztu
2020-2023'' award to the ICC-UB CEX2019-000918-M, ``Unit of Excellence
Maria de Maeztu 2021-2025'' award to ICE CEX2020-001058-M, grant IFT
``Centro de Excelencia Severo Ochoa'' CEX2020-001007-S funded by
MCIN/AEI/\allowbreak 10.13039/\allowbreak 501100011033, as well as
from grants 2021-SGR-249 and 2021-SGR-1526 (Generalitat de Catalunya),
and support from ChETEC-INFRA (EU project no.~101008324).  We also
acknowledge use of the IFT computing facilities.

\appendix

\section{Details of Borexino analysis}
\label{app:borex}

A detailed description of our analysis of the full spectrum of the
Phase-I~\cite{Bellini:2011rx, Bellini:2008mr} and
Phase-II~\cite{Borexino:2017rsf} of Borexino can be found in
Ref.~\cite{Gonzalez-Garcia:2009dpj} and Ref.~\cite{Coloma:2022umy}
respectively.
Here we document the details of our analysis of the Borexino Phase-III
data collected from January 2017 to October 2021, corresponding to a
total exposure of $\text{1431.6 days} \times \text{71.3 tons}$, which
we perform following closely the details presented by the
collaboration in Refs.~\cite{BOREXINO:2020aww, BOREXINO:2022abl}.

\subsection{Analysis of Borexino Phase-III spectrum}
\label{sec:BXIIIspec}

In our fit we use the Borexino spectral data as a function of the
detected hits ($N_h$) on the detector photomultipliers (including
multiple hits on the same photomultiplier) as estimator of the recoil
energy of the electron.  At it was the case in Phase-II, Borexino
divide their Phase-III data in two samples: one enriched (tagged) and
one depleted (subtracted) in the \Nuc[11]{C} events.  The tagged
sample picks up about $36\%$ of the solar neutrino events, while the
subtracted sample accounts for the remaining $64\%$.  In what follows
we denote by $s=$''tagged'' or $s=$''subtracted'' each of the two
samples.  The data and best fit components for the spectrum of the
subtracted sample are shown in Fig.~2(a) of
Ref.~\cite{BOREXINO:2022abl}.  The data points for this sample can
also be found in the data release material in Ref.~\cite{borexdata}.
The corresponding information for the tagged sample was kindly
provided to us by the Borexino collaboration~\cite{borextag}.

The number of expected events $T^0_{s,i}$ in some bin $i$ of data
sample $s$ is the sum of the neutrino-induced signal and the
background contributions.  The main backgrounds come from radioactive
isotopes in the scintillator \Nuc[11]{C}, \Nuc[210]{Bi}, \Nuc[10]{C},
\Nuc[210]{Po} and \Nuc[85]{Kr}.  The collaboration identifies one
additional background due to residual external backgrounds.  With this
\begin{equation}
  T^0_{s,i} = \sum_f S^f_{s,i} + \sum_c B^c_{s,i}
\end{equation}
where the index $f \in \{ \Nuc[7]{Be}, \Nuc{pep}, \Nuc[13]{N},
\Nuc[15]{0}, \Nuc[17]{F}, \Nuc[8]{B}\}$ runs over the solar fluxes
which contribute in the Borexino-III energy range (see
Fig.~\ref{fig:BXIIIspec}), while the index $c \in \{
\Nuc[11]{C},\allowbreak \Nuc[210]{Po},\allowbreak
\Nuc[210]{Bi},\allowbreak \Nuc[85]{Kr},\allowbreak
\Nuc[10]{C},\allowbreak \Nuc{ext} \}$ runs over the background
components.

We compute the solar neutrino signal from flux $f$ in bin $i$,
$S^f_{s,i}$, as
\begin{equation}
  \label{eq:numev}
  S^f_{s,i} = \int_{N_{h,\text{min}}^i}^{N_{h,\text{max}}^i}
  \int \frac{\dd S^f_s}{\dd T_e}(T_e)\,
  \frac{\dd\mathcal{R}}{\dd N_h}(T_e,N_h)\, \dd T_e\, \dd N_h
\end{equation}
where $\dd S_s^f/\dd T_e$ is the differential distribution of
neutrino-induced events from flux $f$ to sample $s$ as a function of
recoil energy of the scattered electrons ($T_e$)
\begin{equation}
  \frac{\dd S^f_s}{\dd T_e}(T_e)
  = \mathcal{F}_s\, \mathcal{N}_\text{tgt}\,
  \mathcal{T}_\text{run}\, \mathcal{E}_\text{cut}
  \sum_\alpha \int \frac{\dd\phi^f_\nu}{\dd E_\nu}\,
  P_{e\alpha}(E_\nu)\,
  \frac{\dd \sigma^{\text{det}}(\nu_\alpha)}{\dd T_e}\,
  \dd E_\nu \,,
\end{equation}
Here $\mathcal{F}_s = 0.3572$ ($0.6359$) is the fraction for $s =
\text{``tagged''}$ (``subtracted'') signal events,
$\mathcal{N}_\text{tgt}$ is the number of $e^-$ targets
(\textit{i.e.}, the total number of electrons inside the fiducial
volume of the detector, corresponding to 71.3~ton of scintillator),
$\mathcal{T}_\text{run} = 1431.6$ days is the data taking time,
$\mathcal{E}_\text{cut} = 98.5\%$ is the overall efficiency (assumed
to be the same as Phase-II), $P_{e\alpha}(E_\nu)$ is the transition
probability between the flavours $e$ and $\alpha$, and
$\dd\sigma^{\text{det}}(\nu_\alpha)/\dd T_e$ is the flavour dependent
$\nu_\alpha - e^-$ elastic scattering detection cross section.
The calculation of $P_{e\alpha}(E_\nu)$ is based on a fully numerical
approach which takes into account the specific distribution of the
neutrino production point in the solar core for the various solar
neutrino flux components as predicted by the SSMs; some technical
details on our treatment of neutrino propagation in the solar matter
can be found in Appendix A of Ref.~\cite{Coloma:2022umy} and in
Sec.~2.4 of Ref.~\cite{Maltoni:2023cpv}.
In addition Eq.~\eqref{eq:numev} includes the energy resolution
function $\dd\mathcal{R}/\dd T_e$ for the detector which gives the
probability that an event with electron recoil energy $T_e$ yields an
observed number of hits $N_h$.  We assume it follows a Gaussian
distribution
\begin{equation}
  \frac{\dd \mathcal{R}}{\dd N_h}
  = \frac{1}{\sqrt{2\pi}\,\sigma_h(T_e)}
  \exp\bigg[ -\frac{1}{2} \bigg(
    \frac{N_h - \bar N_h(T_e)}{\sigma_h(T_e)}\bigg)^2 \bigg]
\end{equation}
where $\bar N_h$ is the expected value of $N_h$ for a given true
recoil energy $T_e$.  We determine $\bar N_h$ via the calibration
procedure described in Ref.~\cite{Coloma:2022umy}, while $\sigma_h$ is
slightly different from Borexino Phase-II analysis.  Concretely, we
derive a relation between $\bar N_h$ and $\sigma_h( \bar N_h)$ which
is
\begin{equation}
  \label{eq:calib}
  \sigma_h( \bar N_h) = 1.21974 + 1.60121 \sqrt{\bar
    N_h} -0.14859 \bar N_h.
\end{equation}

In what respects the backgrounds, we have read the contribution
$B^c_{s,i}$ for each component $c$ in each bin $i$ and for each data
set $s$ from Fig.~2(a) of Ref.~\cite{BOREXINO:2022abl} as well as the
plot provided to us by the collaboration~\cite{borextag}.  These
figures show the best-fit normalization of the different background
components as obtained by the collaboration, and we take them as our
nominal background predictions.\footnote{One technical detail to
notice is that the data in the tables are more thinly binned (817
bins) than the corresponding figures from which we read the
backgrounds (163 bins).  Given the relatively continuous spectra of
the backgrounds, we have recreated the background content of the 817
bins through interpolation.}

\begin{figure}\centering
  \includegraphics[width=0.95\textwidth]{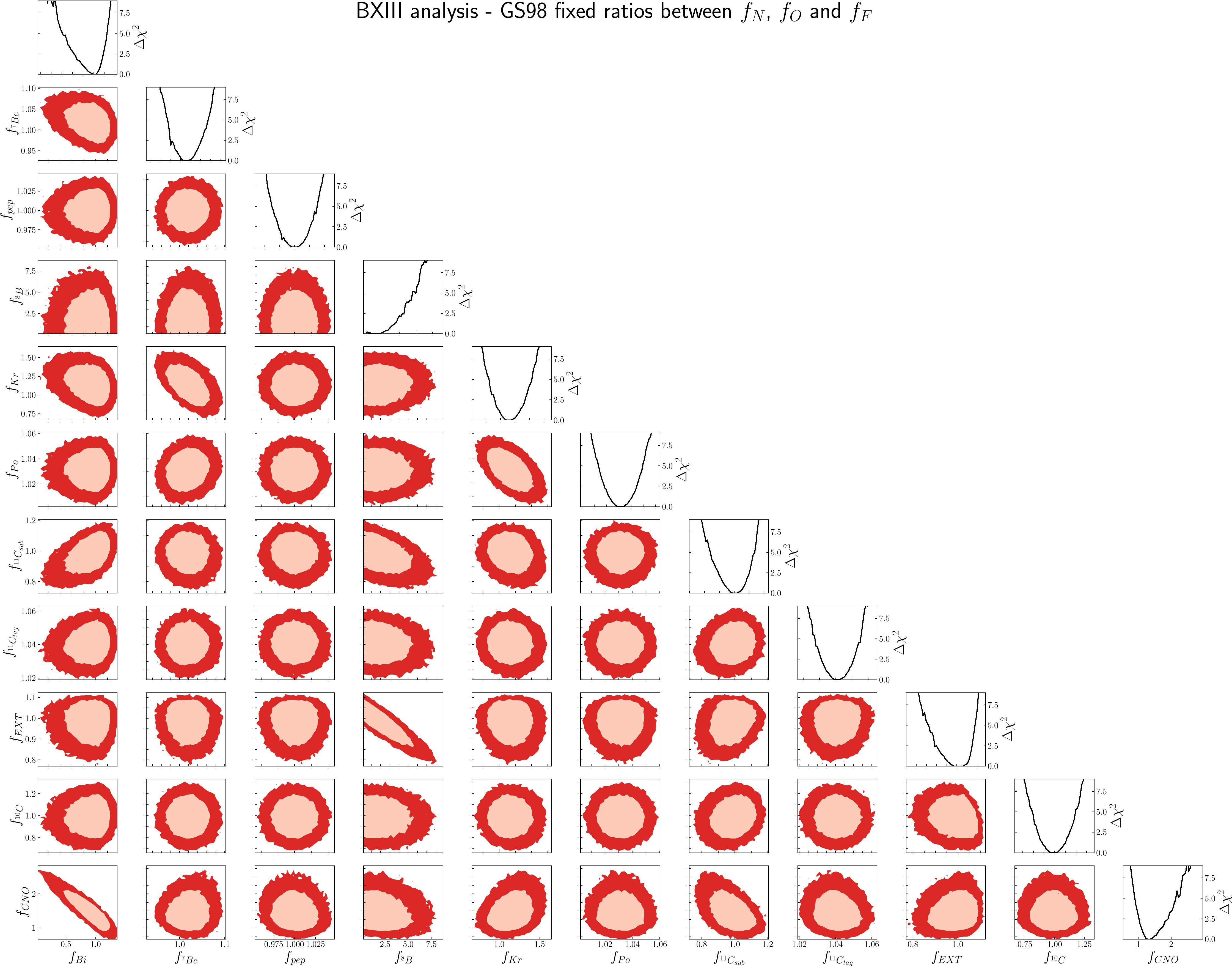}
  \caption{Constraints from our analysis of Borexino Phase-III spectra
    obtained with $\chi^2_\text{BXIII,test}$ in Eq.~\eqref{eq:bxtest}.
    Each panel shows a two-dimensional projection of the allowed
    multi-dimensional parameter space after minimization with respect
    to the undisplayed parameters.  The regions correspond to 90\% and
    99\% CL (2 d.o.f.).  The curves in the right-most panels show the
    marginalized one-dimensional $\Delta\chi^2_\text{BXIII,test}$ for
    each of the parameters.}
  \label{fig:BXIIIfit}
\end{figure}

Our statistical analysis is based on the construction of a $\chi^2$
function built with the described experimental data, neutrino signal
expectations and sources of backgrounds.  Following
Refs.~\cite{BOREXINO:2020aww, BOREXINO:2022abl} we leave the
normalization of all the backgrounds as free parameters with the
exception of \Nuc[210]{Bi}.  The treatment of this background is
paramount to the positive evidence of CNO neutrinos.
As described in~\cite{BOREXINO:2020aww}, the extraction of the CNO
neutrino signal from the Borexino data faces two significant
challenges: the resemblance between spectra of CNO-$\nu$ recoil
electrons and the \Nuc[210]{Bi} $\beta^-$ spectra, and their
pronounced correlation with the \Nuc{pep}-$\nu$ recoil energy
spectrum.  In order to surpass the first challenge, the collaboration
restricted the rate of \Nuc[210]{Bi} for which it sets and upper
limit~\cite{BOREXINO:2022abl}:
\begin{equation}
  R(\Nuc[210]{Bi})\leq
  (10.8\pm 1.0)~\text{cpd} \big/ \text{100\,t} \,,
\end{equation}
while no constraint is imposed on its minimum value which is free to
be as low as allowed by the fit (as long as it remains non-negative).
We will go back to this point in Sec.~\ref{sec:bx3nfit}.
In our analysis we implement this upper limit by constraining the
corresponding normalization factor $f_{\Nuc[210]{Bi}}$ as
\begin{equation}
  f_{\Nuc[210]{Bi}}\leq \bigg( 1 \pm \frac{1.0}{10.8} \bigg),
\end{equation}
With this we construct the $\chi^2_\text{BXIII}$ as
\begin{equation}
  \label{eq:chi2BXIII}
  \chi^2_\text{BXIII}
  = \sum_{s,i} 2 \bigg[ T^0_{s,i} - O_{s,i} + O_{s,i}
  \log\bigg(\frac{O_{s,i}}{T^0_{s,i}} \bigg) \bigg] +
  \bigg( \frac{f_{\Nuc[210]{Bi}} - 1}{\sigma_{\Nuc[210]{Bi}}} \bigg)^2 \,
  \Theta(f_{\Nuc[210]{Bi}} -1)\,,
\end{equation}
where $O_{s,i}$ is the observed number of events in bin $i$ of sample
$s$, and $\sigma_{\Nuc[210]{Bi}} = 1.0 / 10.8$, and $\Theta(x)$ is the
Heaviside step function.

Constructed this way, $\chi^2_\text{BXIII}$ depends on 16 parameters:
the 3 oscillation parameters ($\Dmq_{21}$, $\theta_{12}$,
$\theta_{13}$), 6 solar flux normalizations ($f_{\Nuc[7]{Be}}$,
$f_{\Nuc{pep}}$, $f_{\Nuc[13]{N}}$, $f_{\Nuc[15]{O}}$,
$f_{\Nuc[17]{F}}$, $f_{\Nuc[8]{B}}$) and 7 background normalizations
($f_{\Nuc[210]{Po}}$, $f_{\Nuc[210]{Bi}}$, $f_{\Nuc[85]{Kr}}$,
$f_{\Nuc[10]{C}}$, $f_\text{ext}$ and two different factors
$f_{\Nuc[11]{C}}^\text{tag}$ and $f_{\Nuc[11]{C}}^\text{sub}$ for the
tagged and subtracted samples).

As a first validation of our $\chi^2$ function we perform an analysis
focused at reproducing the results on the solar neutrino fluxes found
by the Borexino collaboration in Ref.~\cite{BOREXINO:2022abl}, and in
particular the positive evidence of CNO neutrinos.  In this test fit
we fix the three oscillation parameters to their best fit value
($\sin^2\theta_{13} = 0.023$, $\sin^2\theta_{12} = 0.307$, $\Dmq_{21}
= 7.5\times 10^{-5}$), and following the procedure of the
collaboration we assume a common normalization factor for the three
CNO fluxes with respect to the SSM ($f_{\Nuc[13]{N}} = f_{\Nuc[15]{O}}
= f_{\Nuc[17]{F}} \equiv f_{\Nuc{CNO}})$.  Furthermore, in order to
break the pronounced correlation with the \Nuc{pep}-$\nu$ recoil
energy spectrum mentioned above, the collaboration introduced a prior
for the \Nuc{pep} neutrino signal flux following the SSM.  Thus we
define
\begin{equation}
  \label{eq:bxtest}
  \chi^2_\text{BXIII,test}=\chi^2_\text{BXIII}
  + \bigg(\frac{f_\text{pep} - 1}{\sigma_{\Nuc{pep}}} \bigg)^2
\end{equation}
with $\sigma_{\Nuc{pep}} = 0.04/2.74$ (for concreteness we choose the
B16-GS98 model for this prior).  The \Nuc[7]{Be} and \Nuc[8]{B} fluxes
are left completely free.

\begin{figure}\centering
  \includegraphics[width=0.6\textwidth]{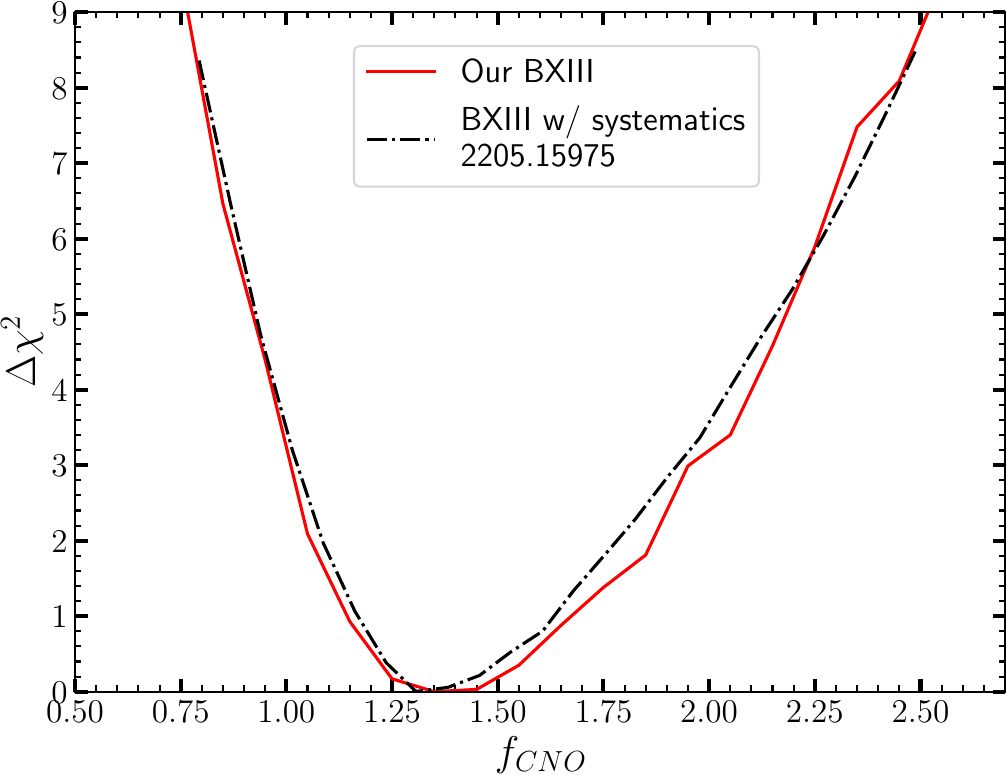}
  \caption{Dependence of $\Delta\chi^2$ of our fit to the Borexino
    Phase-III spectra on the common normalization of the CNO fluxes
    (red line).  For comparison we plot (black dot-dashed line) the
    corresponding results in figure 2(b) of~\cite{BOREXINO:2022abl}
    for their ``Fit w/ Systematics'', divided by the central value of
    the predicted CNO-$\nu$ rate of the B16-GS98 mode.}
  \label{fig:BXIIICNO}
\end{figure}

The results of this 11-parameter fit are shown in
Figs.~\ref{fig:BXIIIfit} and~\ref{fig:BXIIICNO}.  In
Fig.~\ref{fig:BXIIIfit} we plot the allowed ranges and correlations
for the parameters.  Notice that in this figure all parameters are
normalized to the best fit values obtained by the corresponding
analysis of the Borexino collaboration, hence a value of ``1'' means
perfect agreement.  We observe a strong correlation between the
normalization of the CNO fluxes $f_{\Nuc{CNO}}$ and the \Nuc[210]{Bi}
background.  This is expected because, as mentioned before, the
spectrum of CNO neutrinos and that of the \Nuc[210]{Bi} background are
similar (as can also be seen in Fig.~\ref{fig:BXIIIspec} which shows
our best fitted spectra for the two samples).  Still, the two spectra
are different enough so that, under the assumption of the upper bound
on the \Nuc[210]{Bi} background, the degeneracy gets broken enough to
lead to a positive evidence of CNO neutrinos in an amount compatible
with the prediction of the SSMs.

A quantitative comparison with the results of the collaboration is
shown in Fig.~\ref{fig:BXIIICNO} where we plot the dependence of our
marginalized $\Delta\chi^2$ on the common CNO flux normalization,
$f_{\Nuc{CNO}}$, together with that obtained by the collaboration as
extracted from Figure 2(b) of Ref.~\cite{BOREXINO:2022abl} (labeled
``Fit w/ Systematics'' in that figure).\footnote{Figure 2(b) of
Ref.~\cite{BOREXINO:2022abl} shows their $\Delta\chi^2$ as a function
of the CNO-$\nu$ event rate which we divide by the central value of
the expected rate in the B16-GS98 model to obtain the black dot-dashed
curve in Fig.~\ref{fig:BXIIICNO}.}
Altogether, these figures show that our constructed event rates and
the best-fit normalization of the CNO flux reproduce with very good
accuracy those of the fit performed by the collaboration.

\subsection{Allowing free normalizations for the three CNO fluxes}
\label{sec:bx3nfit}

\begin{figure}\centering
  \includegraphics[width=\textwidth]{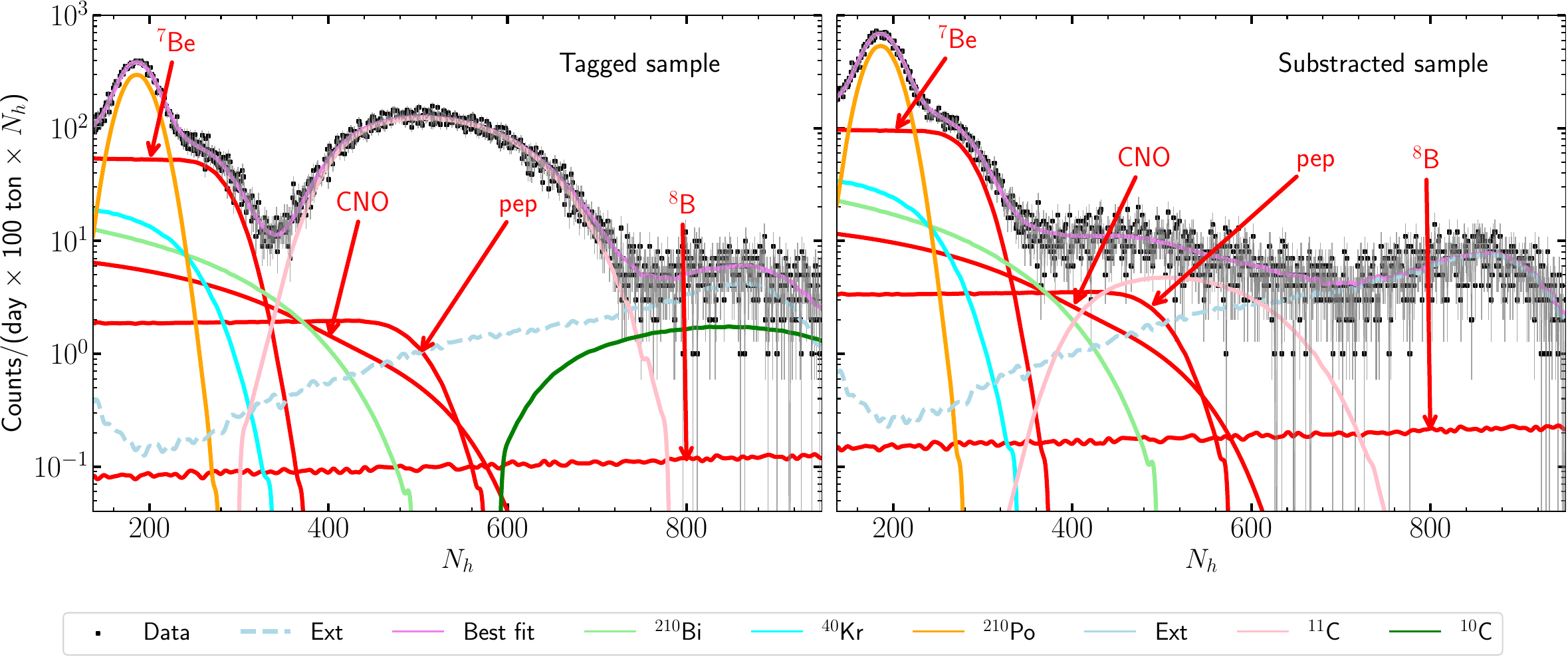}
  \caption{Spectrum for the best-fit normalizations of the different
    components obtained from our fit to the Borexino Phase-III data
    for TFC-tagged (left) and TFC-subtracted (right) events.  In this
    figure ``CNO'' labels the events produced by sum of the three
    fluxes produced in the CNO-cycle, $\Phi_{\Nuc[13]{N}}+
    \Phi_{\Nuc[15]{O}}+ \Phi_{\Nuc[13]{F}}$.}
  \label{fig:BXIIIspec}
\end{figure}

In their analysis of the different phases, the Borexino collaboration
always considers a common shift in the normalization for the three
fluxes of neutrinos produced in the CNO cycle with respect to their
values in the SSM.  On the contrary the normalization of the fluxes
produced in the pp-chain are fitted independently.

\begin{figure}\centering
  \includegraphics[width=0.95\textwidth]{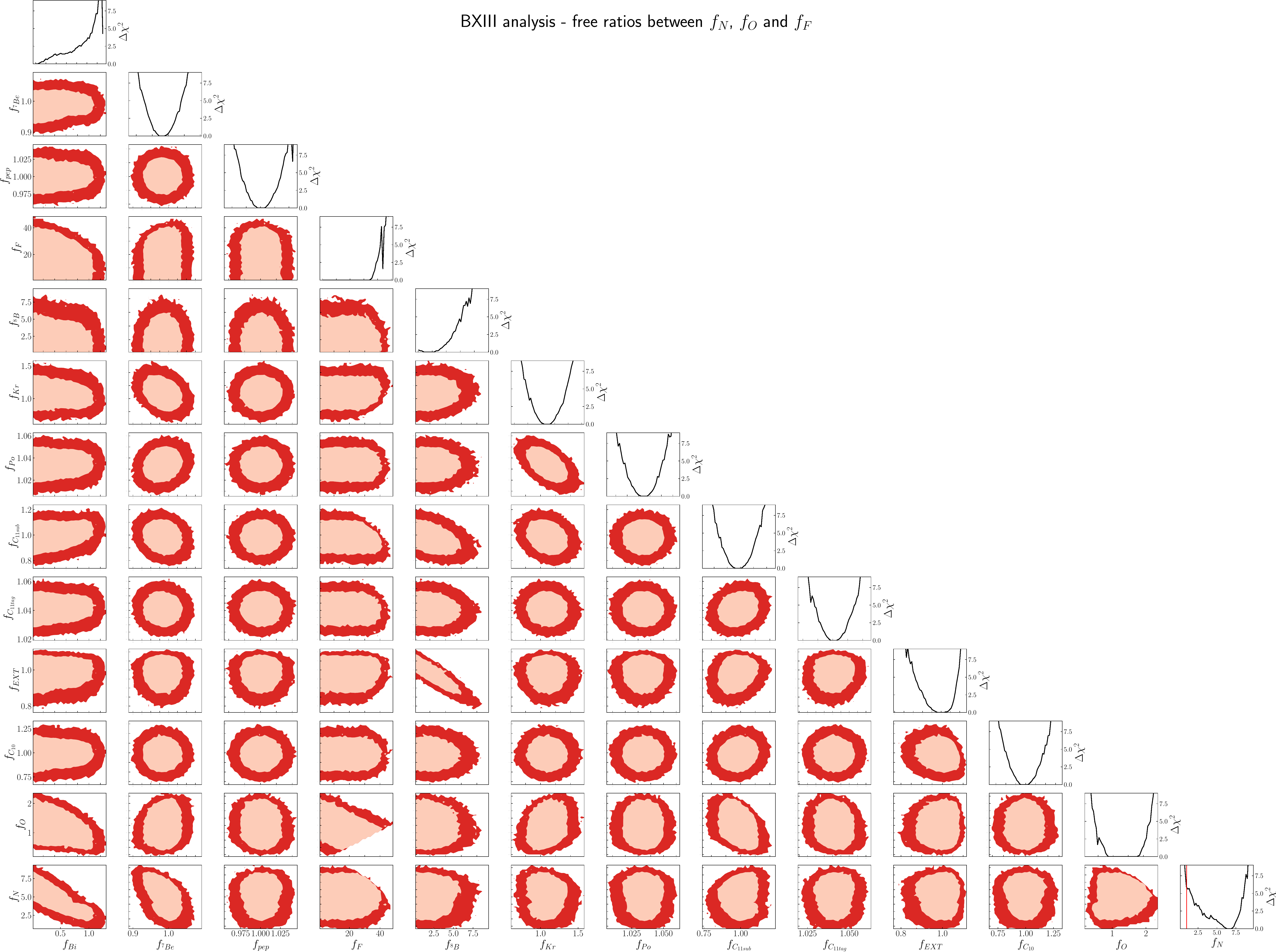}
  \caption{Same as Fig.~\ref{fig:BXIIIfit} but but allowing
    independent variation of the normalizations of the three CNO
    fluxes, only subject to the consistency conditions in
    Eqs.~\eqref{eq:CNOineq1} and~\eqref{eq:CNOineq2}.}
  \label{fig:BXIIINfit}
\end{figure}

In principle, once one departs from the constraints imposed by the
SSM, the normalization of the three CNO fluxes could be shifted
independently, subject only to the minimum set of consistency
relations in Eqs.~\eqref{eq:CNOineq1} and~\eqref{eq:CNOineq2}.  In
fact in our previous works~\cite{Gonzalez-Garcia:2009dpj,
  Bergstrom:2016cbh} we could perform such general analysis.  At the
time there was no evidence of CNO neutrinos and therefore those
analysis resulted into a more general set of upper bounds on their
allowed values compared to those obtained assuming a common shift.
With this as motivation, one can attempt to perform an analysis of the
present BXIII spectra under the same assumption of free normalization.
However, within the present modelling of the backgrounds in the
Borexino analysis, optimized to provide maximum sensitivity to a
positive evidence of CNO neutrinos, such generalized analysis runs
into trouble as we illustrate in Fig.~\ref{fig:BXIIINfit}.  As
expected, allowing three free CNO flux normalizations results in a
weaker constraints on each of the three parameters.  This is
particularly the case for the smaller \Nuc[17]{F} flux which is
allowed to take values as large as $\sim 40$ times the value predicted
by the SSM without however yielding substantial $\chi^2$ improvements
over the standard $f_{\Nuc[17]{F}} = 1$ value.  In the same way
$f_{\Nuc[15]{O}} $ is compatible with the prediction of the SSM,
$\Delta\chi^2(f_{\Nuc[15]{O}} = 1) \simeq 0$, with an upper bound
$f_{\Nuc[15]{O}} \lesssim 2$.\footnote{It is interesting to notice
that the Borexino bound on $\Phi_{\Nuc[17]{F}}$ is about one-half that
on $\Phi_{\Nuc[15]{O}}$.  This is no surprise since the energy spectra
of \Nuc[17]{F} and \Nuc[15]{O} neutrinos are extremely similar hence
neither Borexino nor any other experiment can separate them and what
is actually constrained is their \emph{sum}.  This is reflected in the
clear anticorrelation visible in Fig.~\ref{fig:BXIIINfit}, while the
factor of two stems from the consistency condition in
Eq.~\eqref{eq:CNOineq2}.}  On the contrary the fit results into a
favoured range for \Nuc[13]{N} which, it taken at face value, would
imply an incompatibility with the SSM at large CL:
$\Delta\chi^2(f_{\Nuc[13]{N}}=1) \gtrsim 6$.  This large \Nuc[13]{N}
flux comes at a price of a very low value of the \Nuc[210]{Bi}
normalization, which as seen in the figure is more strongly correlated
with \Nuc[13]{N} than with \Nuc[15]{O} and \Nuc[17]{F}.

\begin{figure}\centering
  \includegraphics[width=\textwidth]{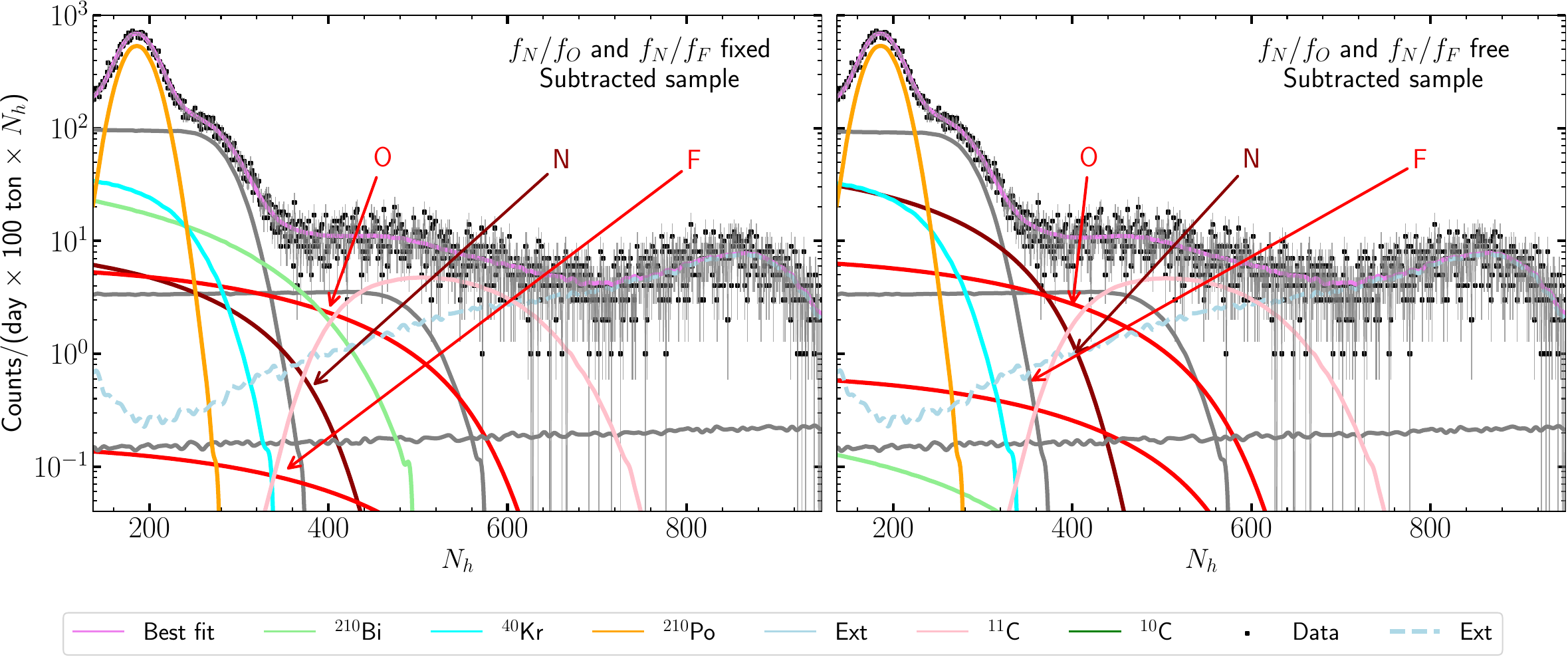}
  \caption{Spectrum of events in the TFC-subtracted sample for the
    best-fit normalizations of the different components obtained from
    two fits to the Borexino Phase-III data.  In the left panel the
    fit is performed assuming a common normalization shift for the 3
    CNO fluxes, while on the right panel the three normalizations are
    allowed to very free subject only the conditions
    Eqs.~\eqref{eq:CNOineq1} and~\eqref{eq:CNOineq2}.}
  \label{fig:subspeccomp}
\end{figure}

To illustrate further this point we show in Fig.~\ref{fig:subspeccomp}
our best fitted spectra of the ``subtracted'' sample for the analysis
where one common normalization for the three CNO fluxes is used (left,
in what follows ``CNO'' fit) and the one where all the three
normalizations are varied independently (right, in what follows ``N''
fit).  Thus the spectra in the left panel of
Fig.~\ref{fig:subspeccomp} are the same as the right panel of
Fig.~\ref{fig:BXIIIspec}, except that now, for convenience, we plot
separately the events from each of the CNO fluxes.  This highlights
clearly the different shape of the spectra of \Nuc[15]{O} and
\Nuc[13]{N}, with \Nuc[15]{O} extending to larger energies.  It is
also evident that \Nuc[13]{N} is the one mostly affected by
degeneracies with the \Nuc[210]{Bi} background.  Comparing the two
panels we see by naked eye that both spectra describe well the data:
in fact, the event rates for \Nuc[15]{O} are comparable in both
panels.  But in the right panel the normalization of the \Nuc[13]{N}
events is considerably enhanced while the \Nuc[210]{Bi} background is
suppressed: this is the option favoured by the fit.

\begin{figure}\centering
  \includegraphics[width=0.59\textwidth]{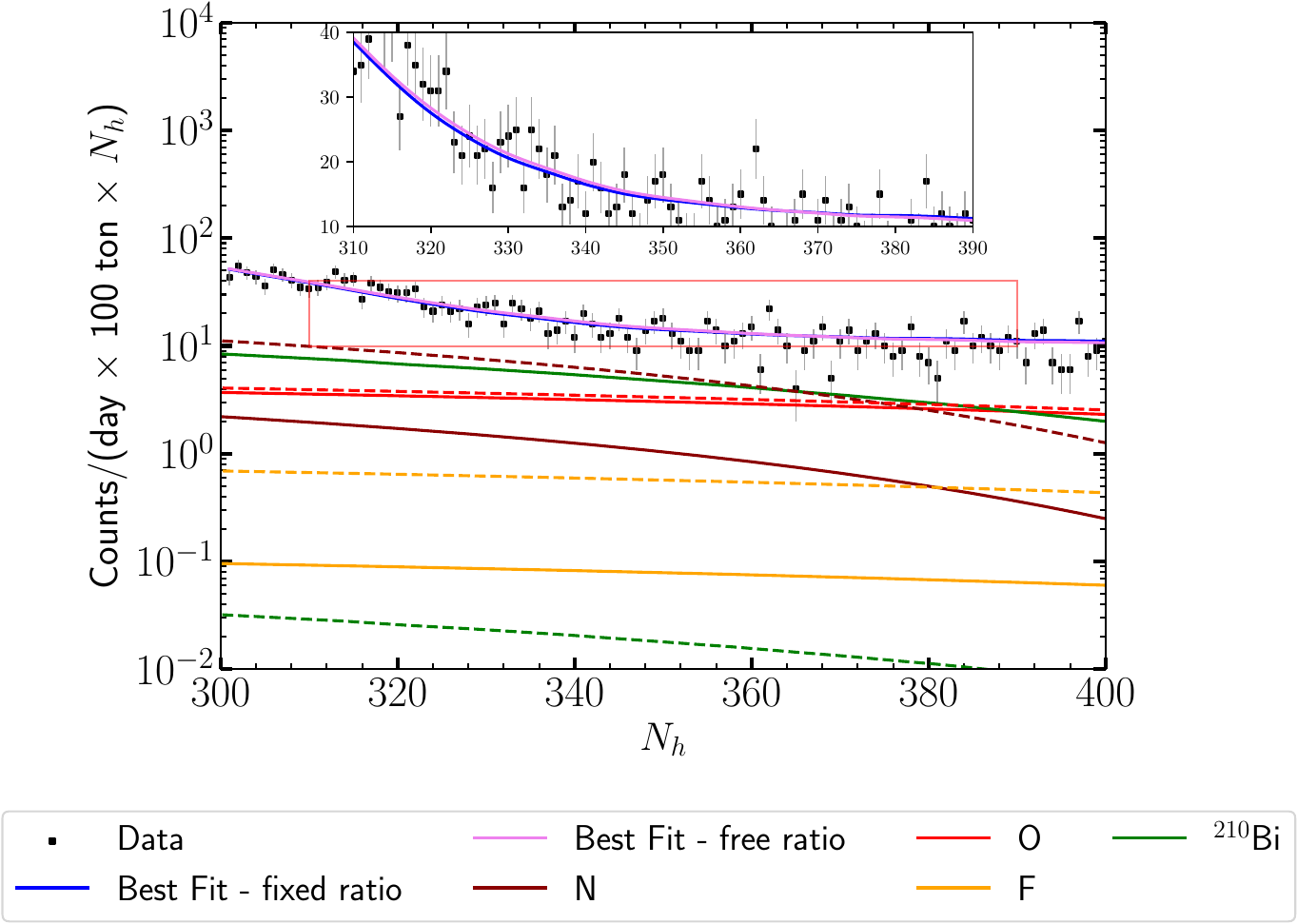}
  \hspace*{-30mm}\hfill
  \raisebox{12mm}{\includegraphics[width=0.455\textwidth]{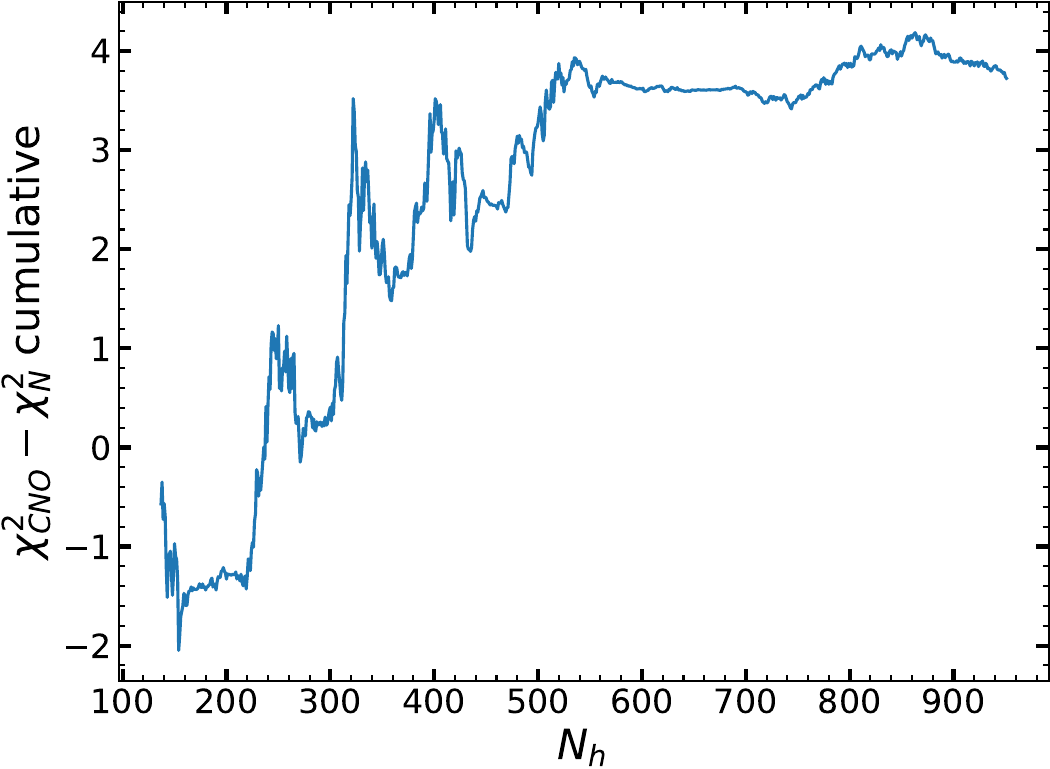}}
  \caption{Left: Spectra of subtracted event rates from best fit CNO
    fluxes and \Nuc[210]{Bi} background in the range of $300 \leq N_h
    \leq 400$ for the ``CNO'' fit (fit with a common normalization
    factor for the three CNO fluxes, full lines) and the ``N'' fit
    (fit with three independent normalizations, dashed lines).  We
    also show the best fit spectra for both fits compared to the data
    as labeled in the figure.  Right: Difference of the value of
    $\chi^2$ in both fits as a function of the maximum $N_h$ bin
    included in the fit.}
  \label{fig:compa2}
\end{figure}

Upon closer examination we find that in the range of $N_h$ spanning
between 300 and 400 photon hits, an increase in the value of
$f_{\Nuc[13]{N}}$ better fits the data while driving
$f_{\Nuc[210]{Bi}}$ towards 0.  In Fig.~\ref{fig:compa2} we show a
blow-up of the spectra in this $N_h$ window.  To quantify the
difference in the quality of the fit for those two solutions and the
relevant range of $N_h$ we plot in the right panel the cumulative
difference of $\chi^2_\text{BXIII,test}$ for the best fits of the
``CNO'' and ``N'' fits as a function of the maximum $N_h$ bin included
in the fit.

Clearly, this anomalously large \Nuc[13]{N} solution is possible only
because the sole information included in the fit for the \Nuc[210]{Bi}
background is the upper bound provided by the collaboration.  Such
upper bound is enough to ensure a lower bound on the amount of CNO
neutrinos, and indeed it results in a positive evidence of CNO fluxes
(in good agreement with at least some of the SSMs) when a common
normalization for the three CNO fluxes is enforced, as reported by the
collaboration in Refs.~\cite{BOREXINO:2020aww, BOREXINO:2022abl} (and
properly reproduced by us, as described in the previous section).  Our
results show that this is the case because the spectrum of
\Nuc[210]{Bi} and \Nuc[15]{O} are sufficiently different.  However,
once the normalization of the three CNO fluxes are not linked
together, the degeneracy between the spectral shape of \Nuc[13]{N} and
\Nuc[210]{Bi} --~together with the lack of a proper estimate for a
\emph{lower} bound on \Nuc[210]{Bi} which is not quantified in
Refs.~\cite{BOREXINO:2020aww, BOREXINO:2022abl}~-- pushes the best-fit
of \Nuc[13]{N} towards unnaturally large values.  In other words, the
background model proposed in Refs.~\cite{BOREXINO:2020aww,
  BOREXINO:2022abl} cannot be reliably employed for fits with
independent \Nuc[13]{N} and \Nuc[15]{O} normalizations.

We finish by noticing that this also implies that the high quality data
of Borexino Phase-III, besides having been able to yield the first
evidence of the presence of the CNO neutrinos, also holds the
potential to discriminate between the contributions from \Nuc[13]{N}
and \Nuc[15]{O}, a potential which may be interesting to explore by
the collaboration.

\subsection{Analysis with Correlated Integrated Directionality Method}

In a very recent work~\cite{Borexino:2023puw} the Borexino
collaboration has presented a combined analysis of their three phases
making use of the Correlated and Integrated Directionality (CID)
method, which aims to enhance the precision of the determination of
the flux of CNO neutrinos.  In a nut-shell, the CID method exploits the
sub-dominant Cherenkov light in the liquid scintillator produced by
the electrons scattered in the neutrino interaction.  These Cherenkov
photons retain information of the original direction of the incident
neutrino, hence they can be used to enhance the discrimination between
the solar neutrino signal and the radioactive backgrounds.

\begin{figure}\centering
  \includegraphics[width=0.554\linewidth]{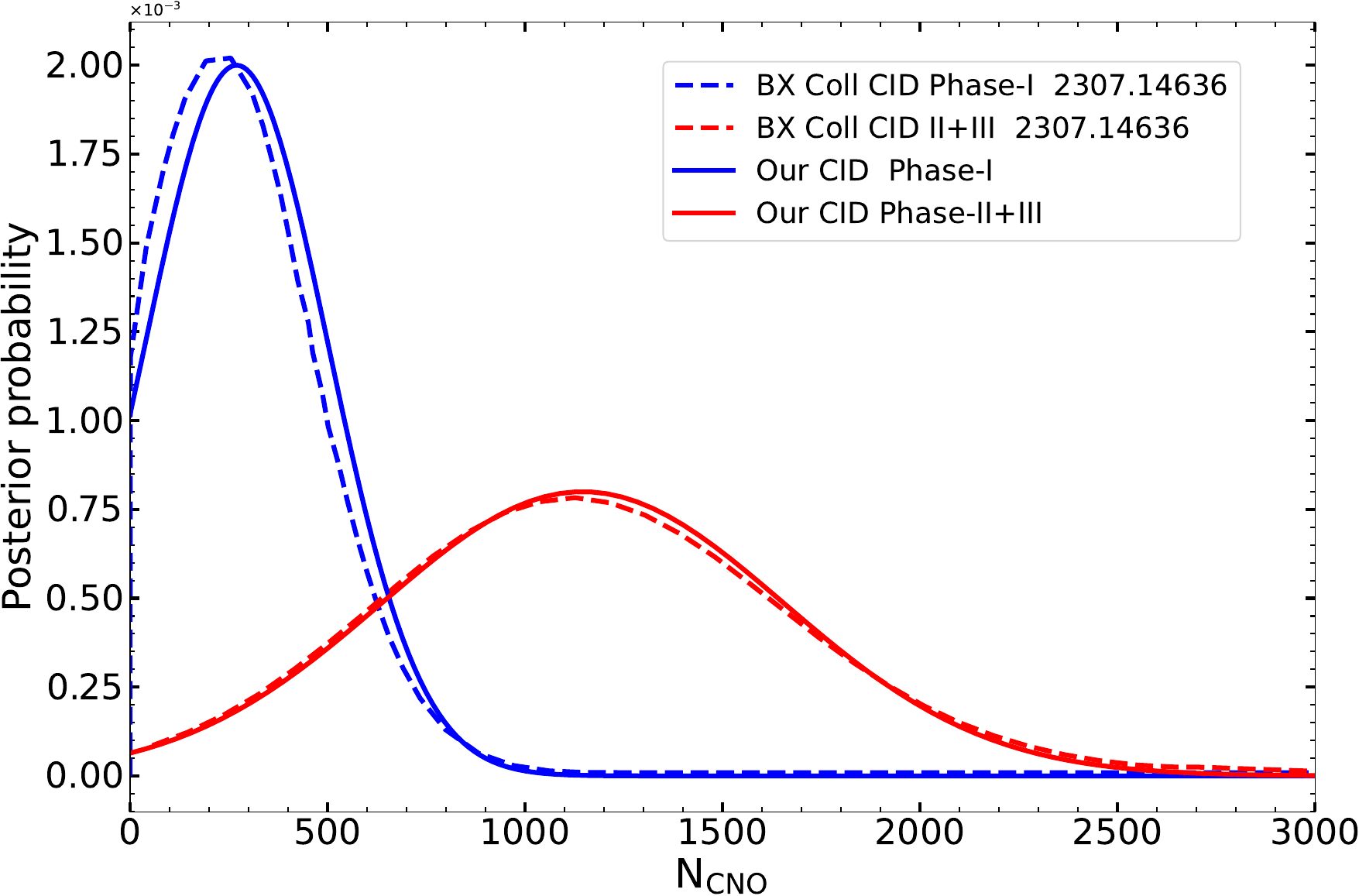}
  \hfill
  \includegraphics[width=0.426\linewidth]{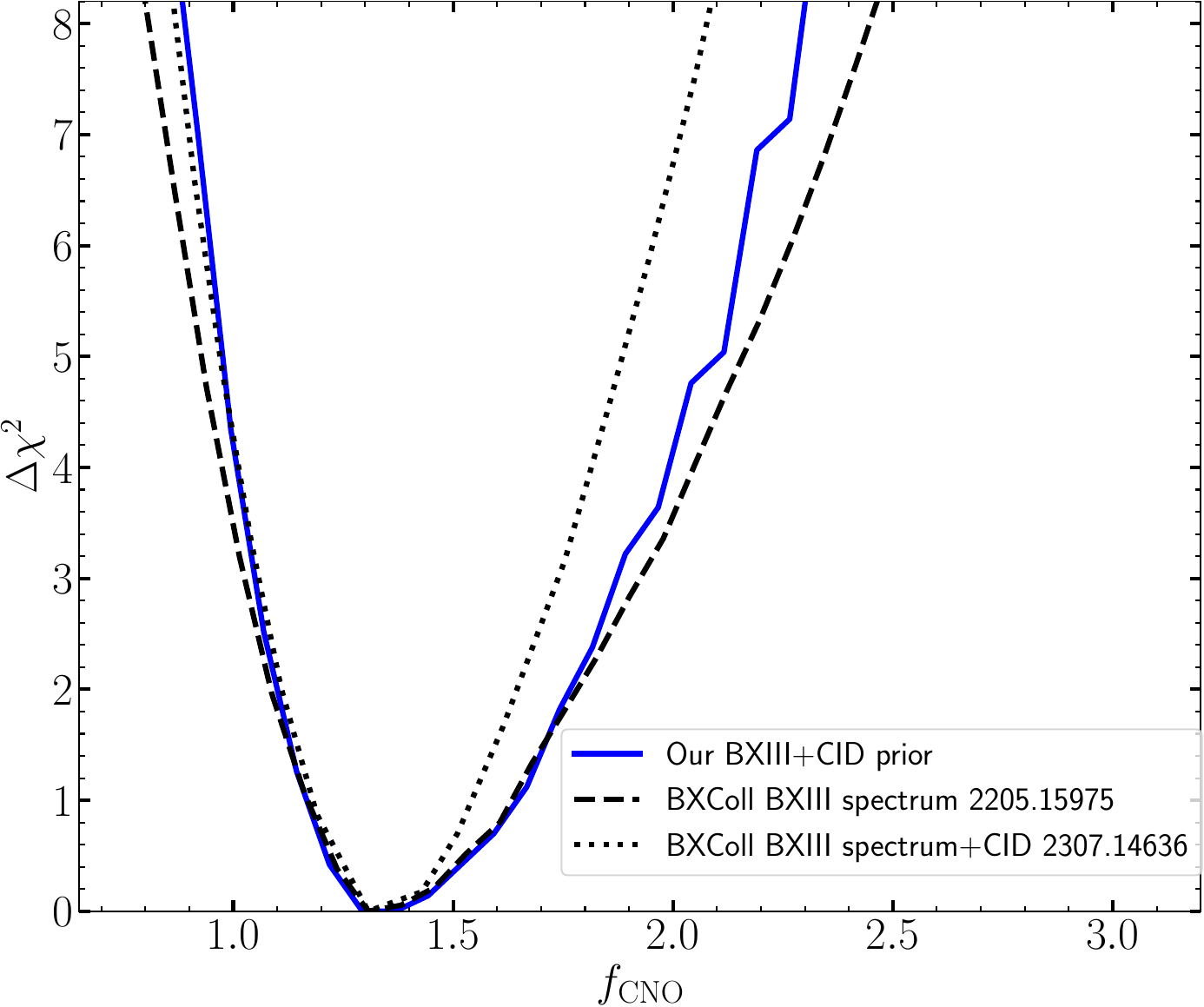}
  \caption{Left: CID posterior probabilities for the number for
    Phase-I and Phase-II+III of CNO-$\nu$ events after constraining
    \Nuc{pep} and \Nuc[8]{B} neutrino events to their SSM expectation.
    Right: Dependence of $\Delta\chi^2$ on the CNO flux normalization
    from our fit to the Borexino Phase-III spectra combined with the
    CID constraint (blue line) compared to that obtained by the
    Borexino collaboration (dotted black line).  For comparison we
    also show the result without CID information (dashed black line).}
  \label{fig:CIDcomp}
\end{figure}

Effectively, the CID analysis results into a determination of the
total number of solar neutrinos detected within a restricted range of
$N_h$ which corresponds to $0.85~\text{MeV} < T_e< 1.3~\text{MeV}$ for
Phase-I and $0.85~\text{MeV} < T_e< 1.29~\text{MeV}$ for Phase-II+III.
In this range the dominant contribution comes from \Nuc{CNO},
\Nuc{pep} and some \Nuc[8]{B}.  The increased fiducial volume for this
analysis brings the exposures to $740.7~\text{days} \times
104.3~\text{ton} \times 55.77\%$ for Phase-I and $2888.0~\text{days}
\times 94.4~\text{ton} \times 63.97\%$ for Phase-II+III.  The
resulting number of solar neutrinos detected is
$N^\text{P-I}_\text{obs} =
643^{+235}_{-224}\text{(stat)}^{+37}_{-30}\text{(sys)}$ for Phase-I
and $N^\text{P-II+III}_\text{obs} =
2719^{+518}_{-494}\text{(stat)}^{+85}_{-83}\text{(sys)}$ for
Phase-II+III.  After subtracting the expected SSM contribution from
\Nuc{pep} and \Nuc[8]{B} the Borexino collaboration obtains the
posterior probability distributions for the number of CNO neutrinos
shown in Fig.~9 of Ref.~\cite{Borexino:2023puw} (which we reproduce in
the left panel in Fig.~\ref{fig:CIDcomp}).  Furthermore, since this
new directional information is independent of the spectral
information, the collaboration proceeded to combine these two priors
on $N_{\Nuc{CNO}}$ with the their likelihood for the Borexino Phase-III
spectral analysis.  This resulted in a slightly stronger dependence of
the combined likelihood on the CNO-$\nu$ rate shown in their Fig.~12
(which we reproduce in the right panel in Fig.~\ref{fig:CIDcomp}).

In order to account for the CID information in our analysis we try to
follow as closely as possible the procedure of the collaboration.
With the information provided on the covered energy range and
exposures for the CID analysis, we integrate our computed spectra of
solar neutrino events in each phase to derive the corresponding total
number of expected events in Phase-I and Phase-II+III.  We then
subtract the SSM predictions for \Nuc{pep} and \Nuc[8]{B} neutrinos
from the observed number of events to derive an estimate for \Nuc{CNO}
neutrinos in in Phase-I and Phase-II+III, and construct a simple
Gaussian $\chi^2(N_{\Nuc{CNO}})$ for Phase-Y (Y=I or II+III)
\begin{equation}
  \label{eq:chi2CID}
  \chi^2_\text{CID,P-Y}(N_{\Nuc{CNO}}) = \bigg(
  \frac{N_\text{obs}^\text{P-Y} - N^\text{P-Y}_\text{SSM,pep}
    - N^\text{P-Y}_\text{SSM,\Nuc[8]{B}} - N_{\Nuc{CNO}}}{\sigma_\text{P-Y}}
  \bigg)^2
\end{equation}
where in $\sigma_{P-Y}$ we add in quadrature the symmetrized
statistical and systematic uncertainties in the number of observed
events.

We plot in the left panel in Fig.~\ref{fig:CIDcomp} our inferred
probability distributions $P_\text{P-Y}(N_{\Nuc{CNO}})\propto
\exp[-\chi^2_\text{P-Y}/2]$ compared to those from Borexino in Fig.~9
of Ref.~\cite{Borexino:2023puw}.  As seen in the figure our simple
procedure reproduces rather well the results of the collaboration for
the Phase-II+III but only reasonably for Phase-I.  This may be due to
differences in the reanalysis of the Phase-I data by the collaboration
in the CID analysis compared to their spectral analysis of 2011.  Our
simulations of the Phase-I event rates are tuned to their 2011 and
there is not enough information in Ref.~\cite{Borexino:2023puw} to
deduce what may have changed.  Thus we decide to introduce in our
analysis the CID prior for the Phase-II+III data but not for Phase-I.

We then combine the CID from Phase-II+III and Phase-III spectral
information as
\begin{equation}
  \chi^2_\text{CID+BXIII,test} = \chi^2_\text{BX-III}
  + \bigg( \frac{f_{\Nuc{pep}}-1}{\sigma_{\Nuc{pep}}} \bigg)^2
  + \chi^2_\text{CID,P-II+III} \,.
\end{equation}
A quantitative comparison with the results of the collaboration for
this combined $\text{CID} + \text{Phase-III}$ spectrum analysis is
shown in the right panel of Fig.~\ref{fig:CIDcomp} where we plot the
dependence of our marginalized $\Delta\chi^2$ on the CNO flux
normalization after including the CID information compared to that
obtained by the collaboration in Fig.~12 of
Ref.~\cite{Borexino:2023puw}.  As seen in the figure, we reproduce
well the improved sensitivity for the lower range of the CNO flux
normalization but our constraints are more conservative in the higher
range, though they still represent an improvement over the
spectrum-only analysis.

Altogether, after all these tests and validations we define the
$\chi^2$ for the full Borexino analysis as
\begin{multline}
  \label{eq:chi2bxtot}
  \chi^2_\text{BX}(\vec\omega_\text{osc},\, \vec\omega_\text{flux})
  = \chi^2_\text{BXI}(\vec\omega_\text{osc},\, \vec\omega_\text{flux})
  + \chi^2_\text{BXII}(\vec\omega_\text{osc},\, \vec\omega_\text{flux})
  \\
  + \chi^2_\text{BXIII}(\vec\omega_\text{osc},\, \vec\omega_\text{flux})
  + \chi^2_\text{CID,P-II+III} (\vec\omega_\text{osc},\, \vec\omega_\text{flux}) \,.
\end{multline}
with $\chi^2_\text{BXIII}(\vec\omega_\text{osc}\,
\vec\omega_\text{flux})$ and $\chi^2_\text{CID,P-II+III}
(\vec\omega_\text{osc}, \vec\omega_\text{flux})$ in
Eqs.~\eqref{eq:chi2BXIII} and~\eqref{eq:chi2CID}, respectively.  We
finish by noticing that the inclusion of the CID information is not
enough to break the large degeneracy between the \Nuc[13]{N} and
\Nuc[210]{Bi} contributions to the spectra discussed in the previous
section.

\bibliographystyle{JHEPmod}
\bibliography{references}

\end{document}